\documentclass[onecolumn]{pasj01_arxiv}
\usepackage{comment}
\usepackage{url}
\usepackage{lineno}
\def\nh{NH$_3$\ } 
\def\HII{H\,\emissiontype{II}\ }
\def\kms{km s$^{-1}$}

\draft 
\Received{2022/10/11}
\Accepted{2023/01/19}
\Published{$\langle$publication date$\rangle$}

\begin{document}
\title{{Ammonia mapping observations of the Galactic infrared bubble N49: Three \nh clumps along the molecular filament} }

\author{Mikito \textsc{Kohno}\altaffilmark{1,2}$^{*}$}%
\author{James O. \textsc{Chibueze}\altaffilmark{3,4}}
\author{Ross A. \textsc{Burns}\altaffilmark{5}}
\author{Toshihiro \textsc{Omodaka}\altaffilmark{6}}
\author{Toshihiro \textsc{Handa}\altaffilmark{6,7}}
\author{Takeru \textsc{Murase}\altaffilmark{6}}
\author{Rin I. \textsc{Yamada}\altaffilmark{2}}
\author{Takumi \textsc{Nagayama}\altaffilmark{8}}
\author{Makoto \textsc{Nakano}\altaffilmark{9}}
\author{Kazuyoshi \textsc{Sunada}\altaffilmark{8}}
\author{Kengo \textsc{Tachihara}\altaffilmark{2}}
\author{Yasuo \textsc{Fukui}\altaffilmark{2}}

\altaffiltext{1}{Astronomy Section, Nagoya City Science Museum, 2-17-1 Sakae, Naka-ku, Nagoya, Aichi 460-0008, Japan}
\altaffiltext{2}{Department of Physics, Graduate School of Science, Nagoya University, Furo-cho, Chikusa-ku, Nagoya, Aichi 464-8602, Japan}
\altaffiltext{3}{Centre for Space Research, Potchefstroom campus, North-West University, Potchefstroom 2531, South Africa}
\altaffiltext{4}{Department of Physics and Astronomy, Faculty of Physical Sciences, University of Nigeria, Carver Building, 1 University Road, Nsukka 410001, Nigeria}
\altaffiltext{5}{National Astronomical Observatory of Japan (NAOJ), National Institutes of Natural Sciences (NINS), 2-21-1 Osawa, Mitaka, Tokyo 181-8588, Japan}
\altaffiltext{6}{Graduate School of Science and Engineering, Kagoshima University, 1-21-35 Korimoto, Kagoshima, Kagoshima 890-0065, Japan}
\altaffiltext{7}{Amanogawa Galaxy Astronomy Research Center (AGARC), Kagoshima University, 1-21-35 Korimoto, Kagoshima, Kagoshima 890-0065, Japan}

\altaffiltext{8}{Mizusawa VLBI Observatory, National Astronomical Observatory of Japan, Osawa 2-21-1, Mitaka-shi, Tokyo 181-8588, Japan}
\altaffiltext{9}{Faculty of Science and Technology, Oita University, 700 Dannoharu, Oita, Oita 870-1192, Japan}

\email{mikito.kohno@gmail.com}
\email{kohno@nagoya-p.jp}


\maketitle

\begin{abstract}
{
{We have carried out the \nh $(J,K)=(1,1),(2,2),$ and $(3,3)$ mapping observations toward the Galactic infrared bubble N49 (G28.83$-$0.25) using the Nobeyama 45\,m telescope.
Three \nh clumps (A, B, and C) were discovered along the molecular filament with the radial velocities of $\sim$ 96, 87, and 89 \kms, respectively.} The kinetic temperature derived from the \nh (2,2)/\nh (1,1) shows {$T_{\rm kin} = 27.0 \pm 0.6$ K} enhanced at Clump B in the eastern edge of the bubble, where position coincides with massive young stellar objects (MYSOs) associated with the 6.7 GHz class II methanol maser source. This result shows the dense clump is locally heated by stellar feedback from the embedded MYSOs. The \nh Clump B also exists at the 88 \kms and 95 \kms molecular filament  intersection. {We therefore suggest that the \nh dense gas formation in Clump B can be explained by a filament-filament interaction scenario.}
On the other hand, {\nh Clump A and C at the northern and southern side of the molecular filament might be the sites of spontaneous star formation because these clumps are located $\sim$\,5$-$10\,pc away from the edge of the bubble.} }
\end{abstract}

\section{Introduction}
{Ionized hydrogen (H\,\emissiontype{II}) regions are formed by the ultra-violet radiation from OB-type stars. 
They are known to be sites of triggered star formation from previous works of observations and theories (e.g., \cite{1998ASPC..148..150E}) and are also identified as ``interstellar bubbles" (e.g., \cite{1975ApJ...200L.107C,1977ApJ...218..377W}).}
For example, the Galactic Legacy Infrared Mid-Plane Survey Extraordinaire (GLIMPSE) by the Spitzer space telescope found $\sim 600$ infrared bubbles in the Galactic plane  ($|l| \leq \timeform{65D}, |b| \leq \timeform{1D}$: \cite{2006ApJ...649..759C,2007ApJ...670..428C}). 
Most of them are \HII regions excited by OB-type stars inside the bubbles \citep{2010A&A...523A...6D}. 
Two star formation scenarios regarding bubbles have recently been discussed as ``collect-and-collapse" (\cite{2010A&A...523A...6D}) and ``cloud-cloud collisions" (\cite{2021PASJ...73S...1F}) based on their morphology and distributions of parent clouds.
The former is shock compression by the expansion of \HII regions which triggers second generation star formation at the edge of the bubble (e.g., \cite{1977ApJ...214..725E}; {\cite{1994MNRAS.268..291W}}; \cite{2005A&A...433..565D,2006ApJ...646..240H,2006A&A...446..171Z,2007A&A...472..835Z,2007MNRAS.375.1291D}; {\cite{2012MNRAS.421..408T}}; \cite{2014A&A...569A..36X}; {\cite{2017A&A...605A..35P,2019MNRAS.487.1517L}}; \cite{2021SciA....7.9511L}).
The latter is the two clouds with different sizes colliding, triggering the formation of exciting stars and forming arc-like morphology (e.g., \cite{1992PASJ...44..203H}; {\cite{1994ApJ...429L..77H}}; \cite{2014AJ....147..141H,2015ApJ...806....7T,2016ApJ...833...85B,2018PASJ...70S..46F,2018PASJ...70S..47O}; {\cite{2019PASJ...71..113K}}; \cite{2019ApJ...872...49F,2021PASJ...73S.338K}; {\cite{2021PASJ...73..880Y,2022ApJ...931..155E}}). 
These scenarios are possibly related to infrared bubbles, however, the dominant mechanism remains unclear.

\begin{figure*}[ht]
\begin{center} 
\includegraphics[width=18cm]{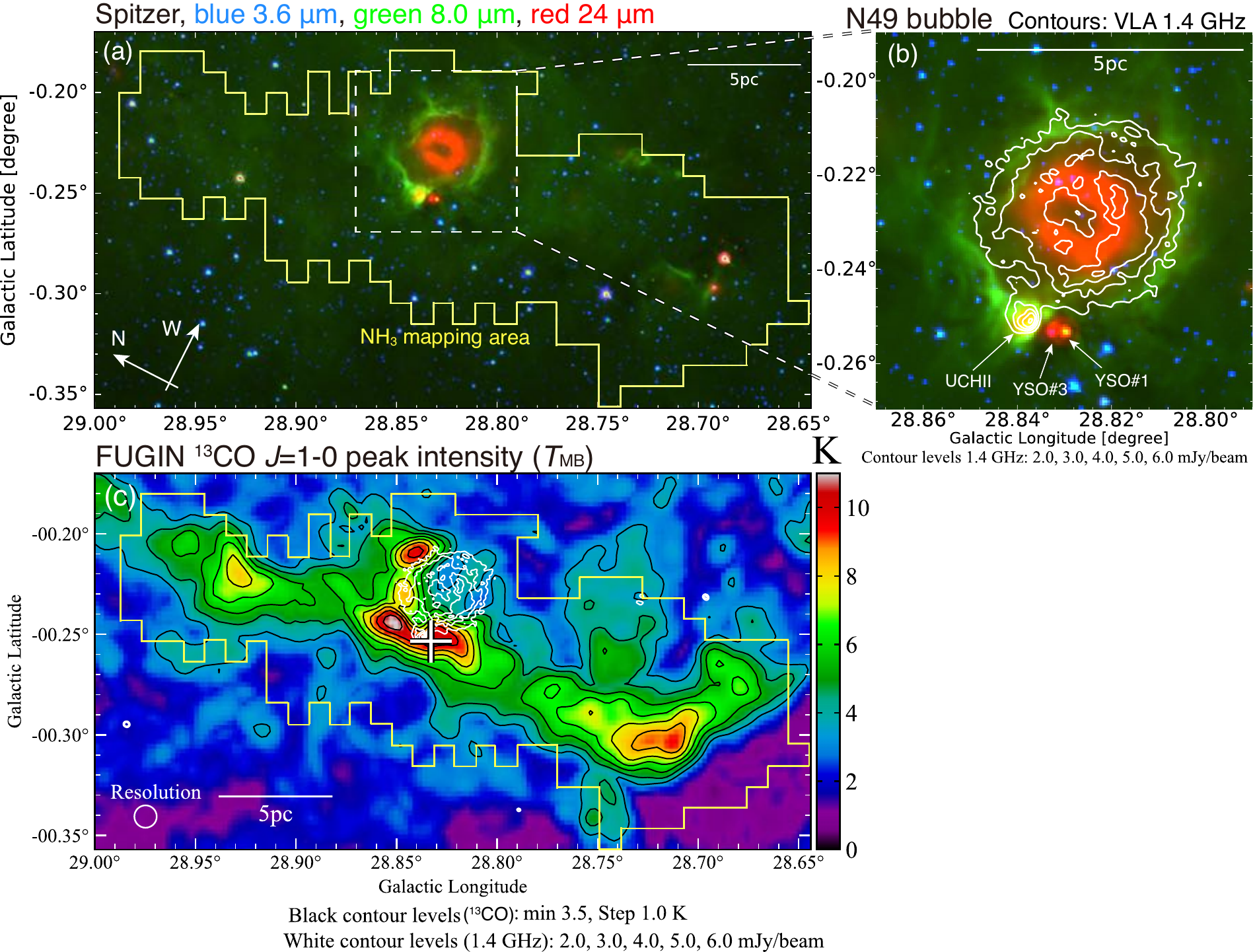}
\end{center}
\caption{(a) Spitzer three color composite image of the infrared bubble N49. Blue, green, and red show $3.6\ \mu$m, $8.0\ \mu$m, and $24\ \mu$m, respectively. Yellow enclosed area indicates the \nh mapping region of our project.  (b) The close-up image of (a). {White contours show the VLA 1.4 GHz continuum image \citep{2006AJ....131.2525H}}. The lowest contour level and interval are 2 and 1 mJy/beam, respectively.
The ultra-compact \HII region, YSO \#1, and YSO \#3 are labeled (see Figure 18 in \cite{2010A&A...523A...6D} and \cite{2008ApJ...681.1341W}).  (c) The peak intensity ({$T_{\rm MB}$}) map of $^{13}$CO $J=$1-0 obtained by the FUGIN project \citep{2017PASJ...69...78U,2019PASJ...71S...2T}. The black lowest contour level and interval are 3.5 and 1.0 K, respectively. White contours show the VLA 1.4 GHz continuum image. The white contour levels are 2.0, 3.0, 4.0, 5.0, and 6.0 mJy/beam. The white cross indicates the position of the 6.7 GHz class II methanol maser source \citep{2008AJ....136.2391C,2009ApJ...702.1615C}.}
\label{bubble}
\end{figure*}

N49 (also known as G28.83$-$0.25) is an infrared bubble at $(l,b) \sim (\timeform{28.83D}, \timeform{-0.23D})$ in the Galactic plane \citep{2010A&A...518L.101Z}, which  is identified as a ``closed bubble" based on AKARI 9\,$\mu$m infrared morphology \citep{2016PASJ...68...37H,2019PASJ...71....6H,2020PASJ...72....5H}.
It has been studied as a representative ``wind-blown bubble" \citep{2010ApJ...713..592E}.
The distance to N49 from the solar system is 5.07 kpc associated with the Scutum Arm \citep{2012AJ....144..173D,2017ApJ...851..140D}.
Figure \ref{bubble}(a) shows the three-color composite image of the Spitzer space telescope \citep{2004ApJS..154....1W}. 
Blue, green, and red present the 3.6 $\mu$m, 8.0 $\mu$m, and 24 $\mu$m images, respectively.
These bands trace the thermal emission from the stars, polycyclic aromatic hydrocarbons (PAH) emission in the photo-dissociation region (e.g., \cite{2007ApJ...657..810D, 2003ARA&A..41..241D}), and hot dust heated by OB-type stars (e.g., \cite{2009PASP..121...76C}), respectively. 
Figure \ref{bubble}(b) presents the close-up image of the N49 bubble. White contours indicate the Very Large Array (VLA) 1.4 GHz continuum image, which traces distributions of the ionized gas. The distribution of the ionized gas coincide the hot dust emission traced by Spitzer 24 $\mu$m image.
YSO $\#$1, YSO $\#$3, and the ultra-compact H II region (UC\HII) exist at the bubble's eastern edge (see Figure 18 in \cite{2010A&A...523A...6D}). 
Previous studies reported that YSO $\#$3 is associated with the 6.7 GHz class II methanol maser embedded in the massive YSOs and referred to as extended green objects (EGOs: \cite{2008AJ....136.2391C,2009ApJ...702.1615C}, and see also data base of astrophysical masers \cite{2019AJ....158..233L} \footnote{\url{https://maserdb.net/object.pl?object=G28.833-0.252}}).
Figure \ref{bubble}(c) shows the $^{13}$CO $J=$1-0 peak intensity map obtained by the Nobeyama 45 m telescope \citep{2017PASJ...69...78U,2019PASJ...71S...2T}. {The optically thinner $^{13}$CO $J=$1-0 can trace a molecular hydrogen density of $n(\rm {H_2}) \sim 10^3$ cm$^{-3}$ (e.g., \cite{1979ApJ...232L..89S,1998AJ....116..336N,1998ApJS..117..387K, 2006ApJS..163..145J, 2019PASJ...71S...2T}). Thus, it is appropriate to study the large-scale structure in the surrounding of the dense gas.
$^{13}$CO has the filamentary structure and arc-like morphology around the ionized gas} at the eastern side of the N49 bubble (see also Figure 12 in \cite{2016AJ....152..117Y}).

{Previous studies of the N49 bubble suggested triggering of star formation} by expanding \HII region \citep{2008ApJ...681.1341W,2012AJ....144..173D}.
\citet{2019RAA....19..183X} performed CO observations toward the molecular filament and bubble. The authors argue that feedback from the \HII region affects star formation and helps to drive the turbulence in the filament.
On the other hand, \citet{2017ApJ...851..140D} found two different radial velocity filaments associated with the bubble. They argued that the filament-filament collision scenario could explain the massive star formation at the edge of the bubble.

Star formation scenarios are discussed in previous works, while the dense gas formation mechanism related to star formation of the N49 bubble and the molecular filament is not yet clear.
{The \nh emission can trace dense gas ($n(\rm {H_2}) \gtrsim 10^4$ cm$^{-3}$) and derive the kinetic temperature of molecular gas because of the excitation of the \nh inversion transitions in the lowest metastable rotational energy states at low temperatures.
It also allows us to derive the optical depth and column density, using the splitting of the inversion transitions into different hyperfine structure components around 23GHz (e.g., \cite{1968PhRvL..21.1701C,1969ApJ...157L..13C,1973ApJ...186..501M, 1983A&A...122..164W,1983ARA&A..21..239H}).} {Previous studies of \nh observations toward the N49 bubble focused only on the dust clumps identified by} the APEX Telescope Large Area Survey of the Galaxy (ATLASGAL: e.g.,\cite{2012A&A...544A.146W}). 
{In order to trace the large-scale structure of the environment around the dust clumps, we performed mapping observations (of about 30 pc) of the \nh inversion transition} of {$(J,K) = (1,1),(2,2),$ and $(3,3)$ using the Nobeyama 45 m telescope (see the yellow enclosed area in Figure \ref{bubble}a and c).
$J$ and $K$ are the quantum number of the total angular momentum and their projection along the molecular axis \citep{1983ARA&A..21..239H}.}
In this paper, we adopted the energy levels of each \nh transition from the Jet Propulsion Laboratory (JPL) spectral line catalog\footnote{\url{https://spec.jpl.nasa.gov}} \citep{1998JQSRT..60..883P}.
This paper is structured as follows: section 2 introduces observations and the archive data; section 3 presents the \nh results; in section 4, we discuss dense gas and star formation scenarios in three \nh clumps along the molecular filament, and in section 5, we show a summary of this paper.

\section{Observations}

\begin{table*}[ht]
{
\tbl{Properties of our observations.}{
\begin{tabular}{cccccccccc}
\hline
\multicolumn{1}{c}{Telescope} & Molecule & Transition & $E_u/k_{\rm B}$ & $\nu_0$  &Receiver & Grid &HPBW  &  Velocity & RMS noise \\
& && [K] &[GHz] &&spacing& &Resolution  & level \\
 (1) & (2) & (3) &(4)& (5) & (6) & (7) &(8) &(9) & (10)\\
\hline
Nobeyama 45 m & NH$_3$ & $(J,K)=(1,1)$& 23.3 & 23.694  &H22 & \timeform{37.5"}  & \timeform{75"}  & 0.39 $\>$km s$^{-1}$ & {$\sim 0.03$ K} \\
 & NH$_3$ &$(J,K)=(2,2)$ & 64.4 & 23.723 & H22 & \timeform{37.5"} & \timeform{75"}  &  0.39 $\>$km s$^{-1}$& {$\sim 0.03$ K}  \\
& NH$_3$ &$(J,K)=(3,3)$ & 123.5 & 23.870 &H22 & \timeform{37.5"} & \timeform{75"}  &  0.39 $\>$km s$^{-1}$& {$\sim 0.03$ K}  \\
\hline
\end{tabular}}\label{obs_param}
\begin{tabnote}
Columns: (1) Telescope name (2) Molecules (3) Transitions. $J$ and $K$ are the quantum number of total angular momentum and their projection along the molecular axis, respectively. (4) Upper energy levels ($E_u$) of inversion transitions above the {ground} state divided by the Boltzmann constant ($k_{\rm B}$). (5) Rest frequency ($\nu_0$) (6) Receiver name (7) Grid spacing of the multi-ON-
OFF switching observations (8) Half-power beam width (9) Velocity resolution (10) r.m.s noise level of the {$T_{\rm MB}$} scale.
\end{tabnote}
}
\end{table*}

\subsection{The KAGONMA project: NH$_3$ mapping observations using the Nobeyama 45 m telescope}
{We carried out the \nh $(J,K)=(1,1),(2,2),$ and $(3,3)$ mapping observations using the Nobeyama 45 m telescope led by Kagoshima University.
{Our \nh survey project is also called ``KAGONMA"\footnote{{See also conference proceedings by \citet{2006JPhCS..54...42H, 2020IAUS..345..353M, Takeba2022}.}} (Kagoshima galactic object survey with the Nobeyama 45-metre telescope by mapping in ammonia lines) and making use of this survey papers on the Galactic Center and massive star forming regions were published} 
(\cite{2007PASJ...59..869N,2009PASJ...61.1023N,2011PASJ...63.1259T,2013ApJ...762...17C,2017PASJ...69...16N,2019PASJ...71...91B,2021arXiv211113481M,2022S255,hirata2022}}). 
{We utilized the dual-polarization and the multi-ON-OFF switching (three ON points per one OFF point) mode}.
{The observational period was from December 2014 to June 2015 (PI T. Omodaka: BU145001\footnote{\url{https://www.nro.nao.ac.jp/~nro45mrt/html/prop/accept/accept2014.html}}).}
The half-power beam width (HPBW) and grid spacing was \timeform{75"} at 23 GHz and \timeform{37.5"}, respectively. 
We used the High Electron Mobility Transistor (HEMT) receiver named H22 at 22 GHz as the front-end. 
The pointing accuracy was checked within \timeform{5"} observing the H$_2$O maser source OH43.8-0.1
$(\alpha_{\rm J2000}, \delta_{\rm J2000})=(\timeform{19h11m54s},\timeform{+09d35'55"})$ before starting observations.
{The FX-type digital spectrometer was used as the back-end system. It is named Spectral Analysis Machine of the 45 m telescope (SAM 45 :\cite{Proc..2011,2012PASJ...64...29K}).}
The system noise temperature ($T_{\rm sys}$) including the atmosphere was 150-250 K during observations.
We used 8 intermediate frequencies (IFs) to observe the \nh $(J,K)=(1,1), (2,2)$, $(3,3)$, and H$_2$O maser with the dual-polarization simultaneously.
Each IF's frequency bandwidth and resolution were 125 MHz and 30.52 kHz corresponding to the velocity coverage of 1600 \kms and spacing of 0.39 \kms at 23 GHz, respectively.
The chopper wheel method was adopted to calibrate to the intensity of the antenna temperature ($T_a^*$: \cite{1976ApJS...30..247U,1981ApJ...250..341K}). 
We used the NEWSTAR software package \citep{2001ASPC..238..522I}\footnote{\url{https://www.nro.nao.ac.jp/~nro45mrt/html/obs/newstar/index.html}} to make the cube file from the raw data. 
{We subtract baselines with a first-order polynomial function from the spectra. The baseline ranges were set up emission-free channels of each spectrum.}
Then, we applied the hanning smoothing (width=5) in the velocity domain to improve the signal-to-noise ratio.
{The antenna temperature was converted to the main-beam temperature ($T_{\rm MB}$) using the relation of $T_{\rm MB}= T_a^*/\eta_{\rm MB}$. 
The main-beam efficiency ($\eta_{\rm MB}$) is 0.835, the average value of two channels of the H22 receiver in the Nobeyama 45 m telescope.}
{The kinetic temperature and physical parameters are not dependent on the main-beam efficiency because we can derive them from intensity ratios of different transitions around 23 GHz obtained by the same instrument of a radio telescope.}
{The root-mean-square noise level for the final data was $\sim 0.03$ K at the $T_{\rm MB}$ scale (corresponding to $\sim 0.025$ K at the $T_{a}^*$ scale)}. 
{The parameters of our \nh data are summarized in Table \ref{obs_param}.}


\subsection{Archival data}
{We utilized the $^{13}$CO $J=$1-0 data comparing with \nh spatial and velocity distributions. It was obtained by the FOREST Unbiased Galactic plane Imaging survey with the Nobeyama 45-m telescope (FUGIN\footnote{\url{https://nro-fugin.github.io}}: \cite{2017PASJ...69...78U,2018PASJ...70S..50K}; Nishimura et al. 2018,{2021}; \cite{2019PASJ...71S...2T,2019PASJ...71S...1S,2020PASJ...72...43N}; {\cite{2021PASJ...73..568S}}; \cite{2021PASJ...73S.172F,2021PASJ...73S.129K}).
The FUGIN CO survey adopted the On-the-fly (OTF) mapping mode \citep{2008PASJ...60..445S} and used the FOur-beam REceiver System on the 45m Telescope (FOREST) with the four beams, dual-polarization, side-band separating receiver \citep{2016SPIE.9914E..1ZM,2019PASJ...71S..17N}.
{The HPBW and effective resolution convolved by the Gaussian kernel function is  $\timeform{15"}$ and  $\timeform{21"}$  at 110 GHz, respectively.}
The effective velocity resolution is 1.3 \kms. 
{Smoothing the data cube by a Gaussian kernel function resulted in a resolution of $\sim \timeform{40"}$.}
The r.m.s noise level is $\sim 0.3$ K at the main beam temperature scale.}
We used the 1.4 GHz ($=$ 20 cm) radio continuum image with the angular resolution of $\sim$ \timeform{6"} taken from the Multi-Array Galactic Plane Imaging Survey (MAGPIS) with VLA \citep{2005AJ....130..586W,2006AJ....131.2525H}. The 1.4 GHz radio continuum image is useful to investigate the free-free emission corresponding to the ionized gas from \HII regions.
We also obtained the archive data of the Spitzer Space telescope with the wavelength of $3.6\ \mu$m, $8\ \mu$m \citep{2003PASP..115..953B,2009PASP..121..213C}, and $24\ \mu$m \citep{2009PASP..121...76C}. 
The 870 $\mu$m dust continuum image is also taken from ATLASGAL\footnote{\url{http://www3.mpifr-bonn.mpg.de/div/atlasgal/}} \citep{2009A&A...504..415S}.
These fits cubes were taken from the MAGPIS web page\footnote{\url{https://third.ucllnl.org/gps/}}.

\subsection{{Software packages}}
{In this paper, we used the following software packages to create the maps, extract the spectra, and calculate physical parameters (Astropy: \cite{2013A&A...558A..33A, 2018AJ....156..123A}, NumPy: \cite{2011CSE....13b..22V}, Matplotlib: \cite{2007CSE.....9...90H}, IPython: \cite{2007CSE.....9c..21P}, Miriad: \cite{1995ASPC...77..433S}, and APLpy:  \cite{2012ascl.soft08017R}).}

\section{Results}
\subsection{NH$_3$ spatial distributions}
\begin{figure*}[t]
\begin{center} 
\includegraphics[width=13cm]{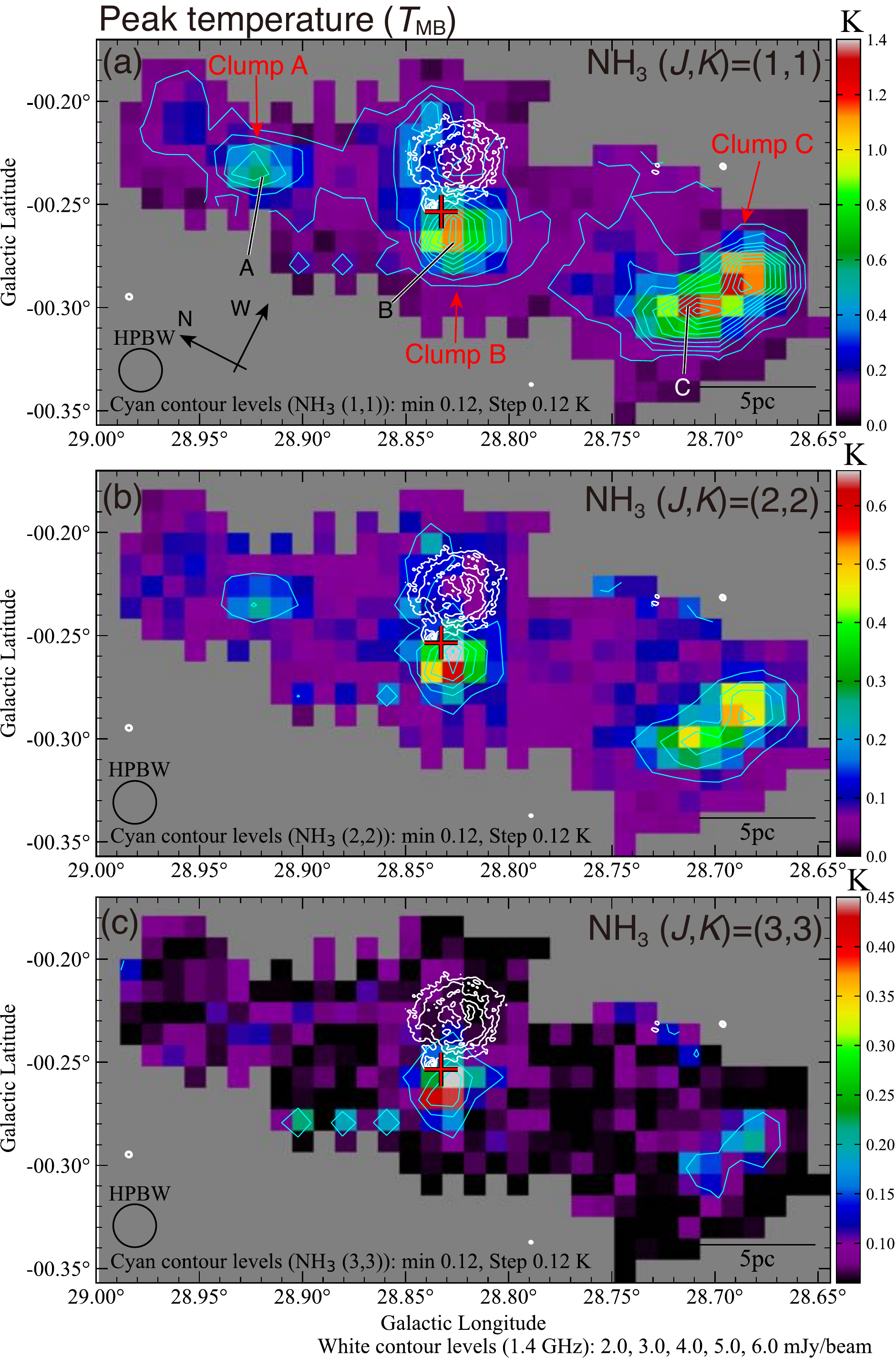}
\end{center}
\caption{Peak intensity ({$T_{\rm MB}$}) map of (a) NH$_3$ $(J,K) = (1,1)$, (b) NH$_3$ (2,2), and (c) NH$_3$ (3,3). The lowest cyan contour level and intervals of \nh are {0.12 K} ($\sim 4\sigma$) and {0.12 K} ($\sim 4\sigma$), respectively. Three \nh clumps are labeled in panel (a). A, B, and C show the positions of the spectra in Figure \ref{spec}. {The white contour levels of the VLA 1.4 GHz continuum image are 2.0, 3.0, 4.0, 5.0, and 6.0 mJy/beam.} Red cross indicates the position of the 6.7 GHz class II methanol maser source \citep{2008AJ....136.2391C,2009ApJ...702.1615C}. }
\label{peakt}
\end{figure*}

Figure \ref{peakt} shows the peak temperature map of the main line in (a) \nh $(J,K) = (1,1)$, (b) \nh (2,2), and (c) \nh (3,3). The white contours present the ionized gas distribution traced by {the VLA 1.4 GHz radio continuum image}. 
{We identified three clumps in the \nh (1,1) line as revealed by the intensity peaks along the molecular filament (Figure \ref{peakt}a).}
Hereafter, we called these clumps ``Clump A", ``Clump B", and ``Clump C".
The \nh (1,1) peak in Clump B exists on the eastern side of the N49 bubble {offset from the ionized gas.} 
We detected the \nh (2,2) in Clump A, B, and C (Figure \ref{peakt}b). 
The \nh (3,3) emission is enhanced in Clump B at the eastern edge of the bubble (Figure \ref{peakt}c).

\clearpage

\begin{figure*}[h]
\begin{center} 
\includegraphics[width=8cm]{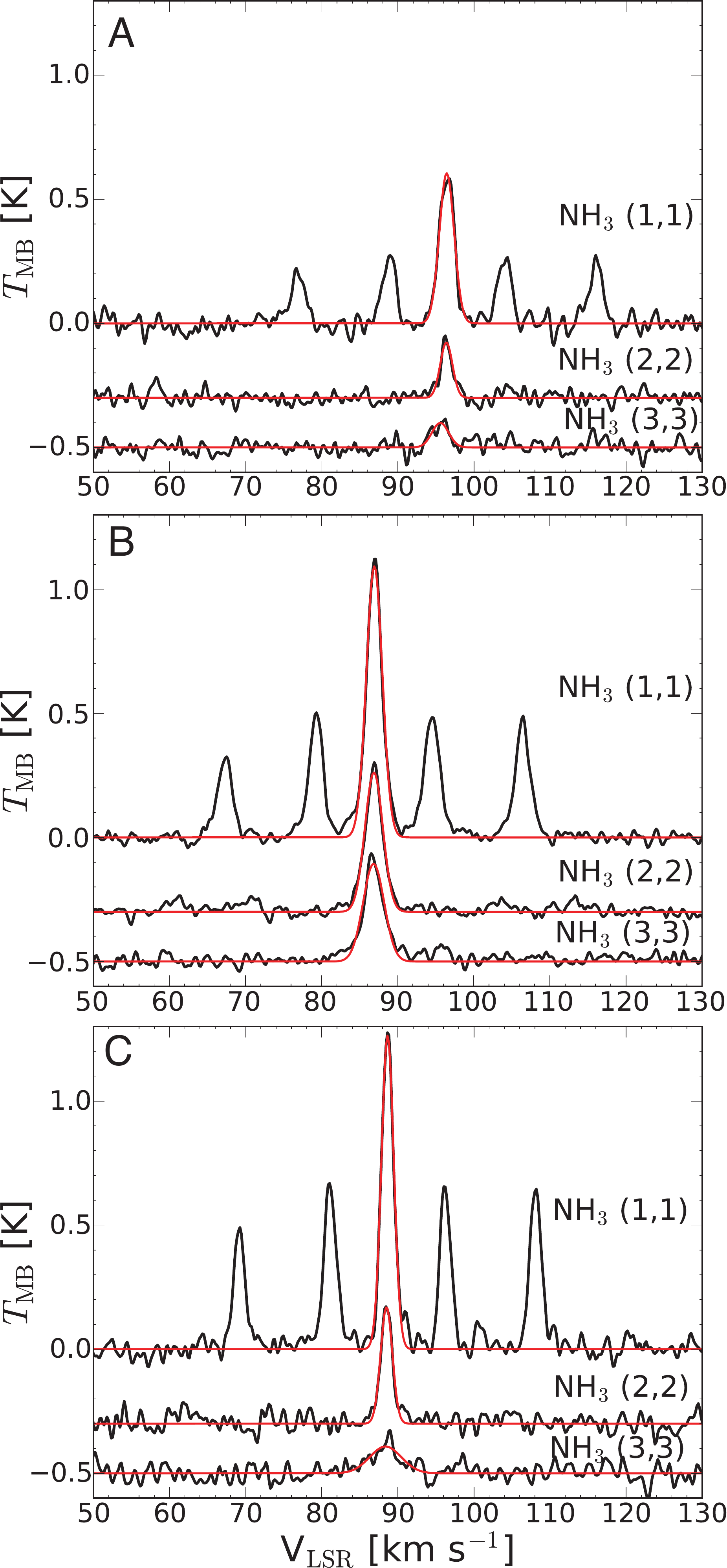}
\end{center}
\caption{NH$_3$ spectra at A $(l,b)=(\timeform{28.921D}, \timeform{-0.237D})$, B $(l,b)=(\timeform{28.827D}, \timeform{-0.268D})$, and C $(l,b)=(\timeform{28.712D}, \timeform{-0.299D})$. These positions are also presented in Figure \ref{peakt}(a). {Red lines show the results from a Gaussian function fitted to the emission at the main component.} \nh (1,1) emissions include four satellite components of $F_1 = 1 \rightarrow 0$, $F_1 =1 \rightarrow 2$,$F_1 =2\rightarrow 1$, and $F_1 =0\rightarrow 1$ (from left to right on the velocity axis).}
\label{spec}
\end{figure*}

Figure \ref{spec} shows the \nh peak spectrum at Clump A, B, and C.
The locations of each spectrum are labeled in Figure \ref{peakt}(a). 
In the \nh (1,1) emission, the main line {exists as the center velocity} of each spectrum.
We also detected the hyper-fine structures of the inner ($F_1 = 1 \rightarrow 2, 2 \rightarrow 1$) and outer satellite lines ($F_1 = 1 \rightarrow 0$, $0 \rightarrow 1$), where 
$F_1$ is the quantum number of the angular momentum, contributing to the nitrogen nuclear-spin. 
The intensity peak of \nh (1,1) exists at Clump C, {while \nh (3,3) has a peak intensity at Clump B}. 
We derived the peak temperature, peak velocity, and full-width half maximum (FWHM: hereafter, we call it the line width) obtained by a Gaussian fitting toward the main line of each profile. 
The profile parameters are summarized in Table \ref{spec_param}. {The errors are the standard deviations of each fitting parameter.}

\begin{table*}[h]
{
\tbl{NH$_3$ spectra parameters of the main line in each \nh clump}{
\begin{tabular}{cccccccccc}
\hline
\multicolumn{1}{c}{Name}& $l$ & $b$ & Transition &${T_{\rm MB}}$ & $v_{\rm LSR}$ &  $\Delta v$\\
&[degree]&[degree] & $(J,K)$ & [K] & [km s$^{-1}$] & [km s$^{-1}$] \\
(1) & (2) & (3) &(4)& (5) & (6) & (7)  \\
\hline
A & $28.921$ & $-0.237$ &(1,1) & ${0.61 \pm 0.02}$ & $96.4\pm 0.04$  & $2.3 \pm 0.10$ \\
& &  &(2,2) & ${0.22 \pm  0.01}$ & $96.4 \pm 0.05$  & $1.9 \pm 0.11$ \\
& &  &(3,3) & ${0.10 \pm 0.01}$ & $95.6 \pm 0.12$  & $2.6 \pm 0.29$ \\
\hline
B & $28.827$& $-0.268$ &(1,1) & ${1.09 \pm 0.04}$ & $86.9 \pm 0.02$  & $2.5 \pm 0.10 $ \\
& &  &(2,2) & ${0.56 \pm 0.007}$ &  $86.9 \pm 0.02$ & $2.8 \pm 0.04$ \\
& &  &(3,3) & ${0.39 \pm 0.007}$ &  $86.8 \pm 0.03$ & $3.3 \pm 0.06$ \\
\hline
C & $28.712$& $-0.299$ &(1,1) & ${1.26 \pm 0.05}$ & $88.7 \pm 0.04$  & $1.9 \pm 0.10$ \\
& &  &(2,2) & ${0.47 \pm 0.01}$ & $88.5 \pm 0.03$  & $1.8 \pm 0.06$\\
& &  &(3,3) & ${0.11 \pm 0.008}$ & $88.4 \pm 0.19$  & $4.9 \pm 0.44$ \\
\hline
\end{tabular}}\label{spec_param}
\begin{tabnote}
Columns: (1) Name of each spectra (2) Galactic Longitude (3) Galactic Latitude (4) Rotational transition (5) {Peak temperature ($T_{\rm MB}$)} (6) Peak velocity by a single Gaussian fitting (7) The line width ($2 \sigma \sqrt{2 \ln 2}$) of spectra by a Gaussian fitting, where $\sigma$ is the standard deviation of the line profile. {The errors are the standard deviations of each fitting parameter.}
\end{tabnote}
}
\end{table*}

\subsection{Three \nh clumps with the different radial velocity}
\begin{figure*}[h]
\begin{center} 
\includegraphics[width=16.5cm]{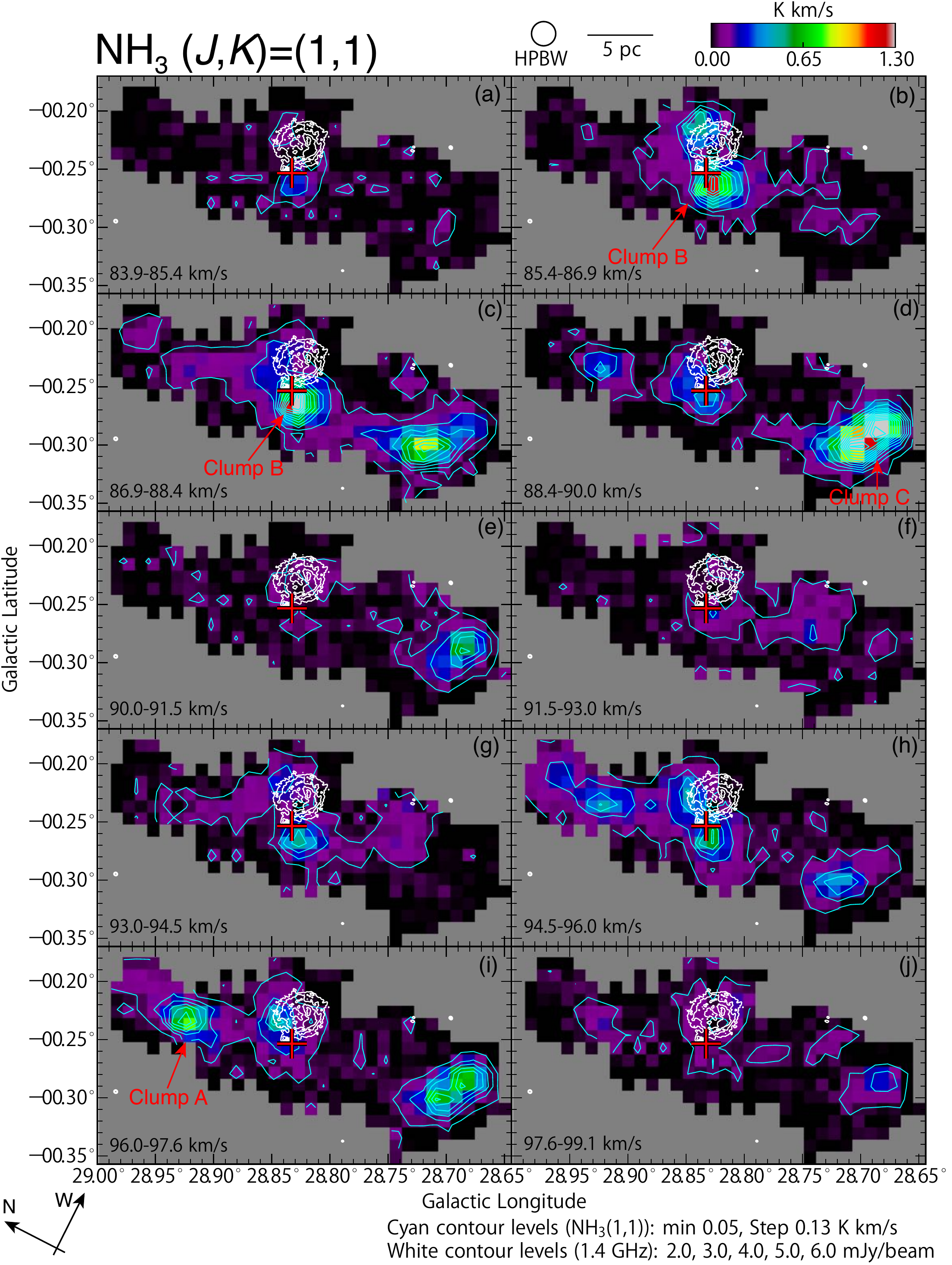}
\end{center}
\caption{{Velocity channel map of NH$_3$ $(J,K)=(1,1)$. The lowest cyan contour level and intervals are {0.05 K km s$^{-1}$ ($\sim 3\sigma$) and 0.13 K km s$^{-1}$} ($\sim 8\sigma$), respectively. The red cross and white contours are {the same as in Figure \ref{peakt}a.} {The white contour levels of the VLA 1.4 GHz continuum image are 2.0, 3.0, 4.0, 5.0, and 6.0 mJy/beam.}}}
\label{ch}
\end{figure*}

\begin{figure*}[h]
\begin{center} 
\includegraphics[width=16.5cm]{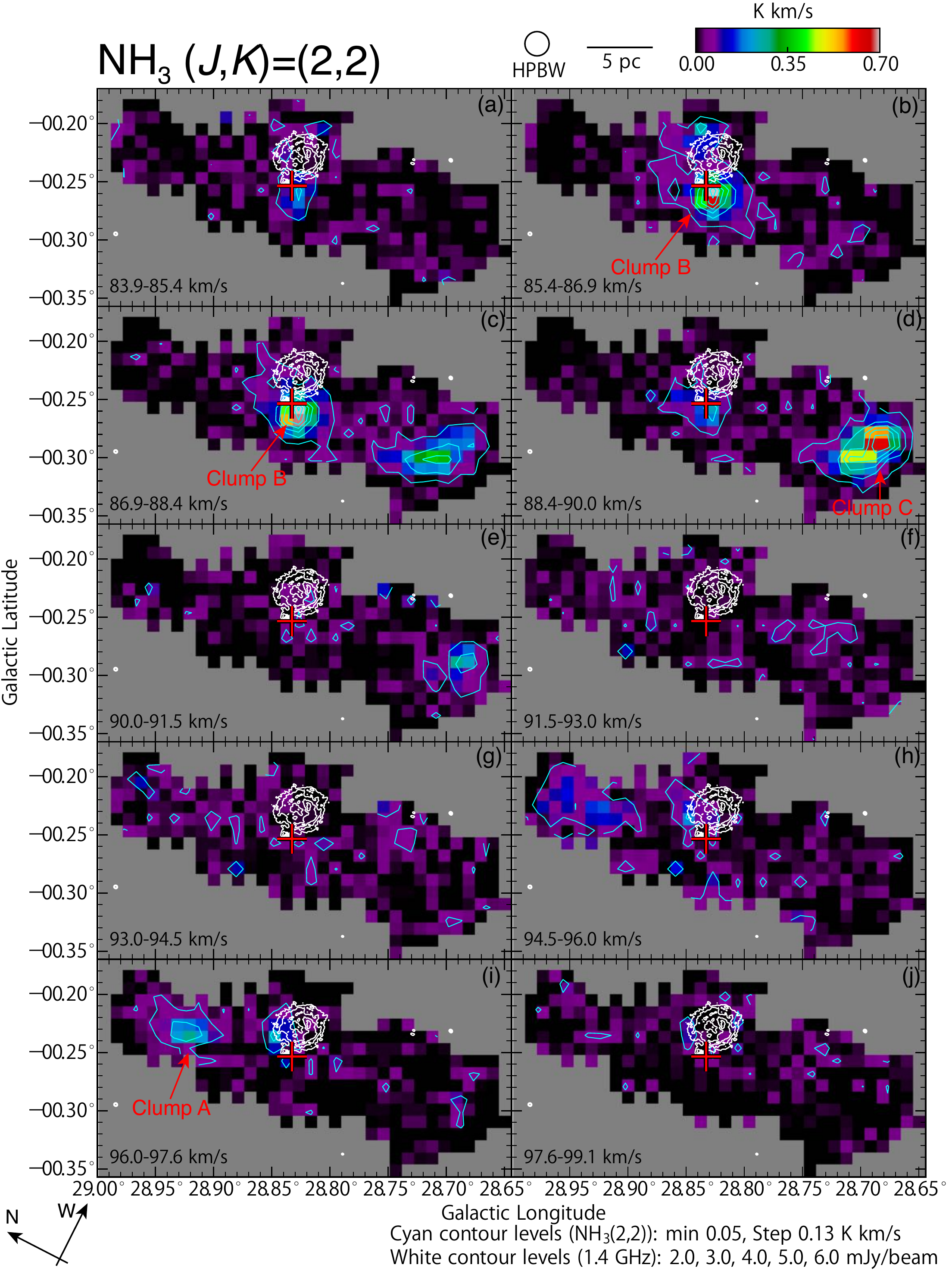}
\end{center}
\caption{Same as Figure \ref{ch}, but for NH$_3$ $(J,K)=(2,2)$.}
\label{ch2}
\end{figure*}

\begin{figure*}[h]
\begin{center} 
\includegraphics[width=16.5cm]{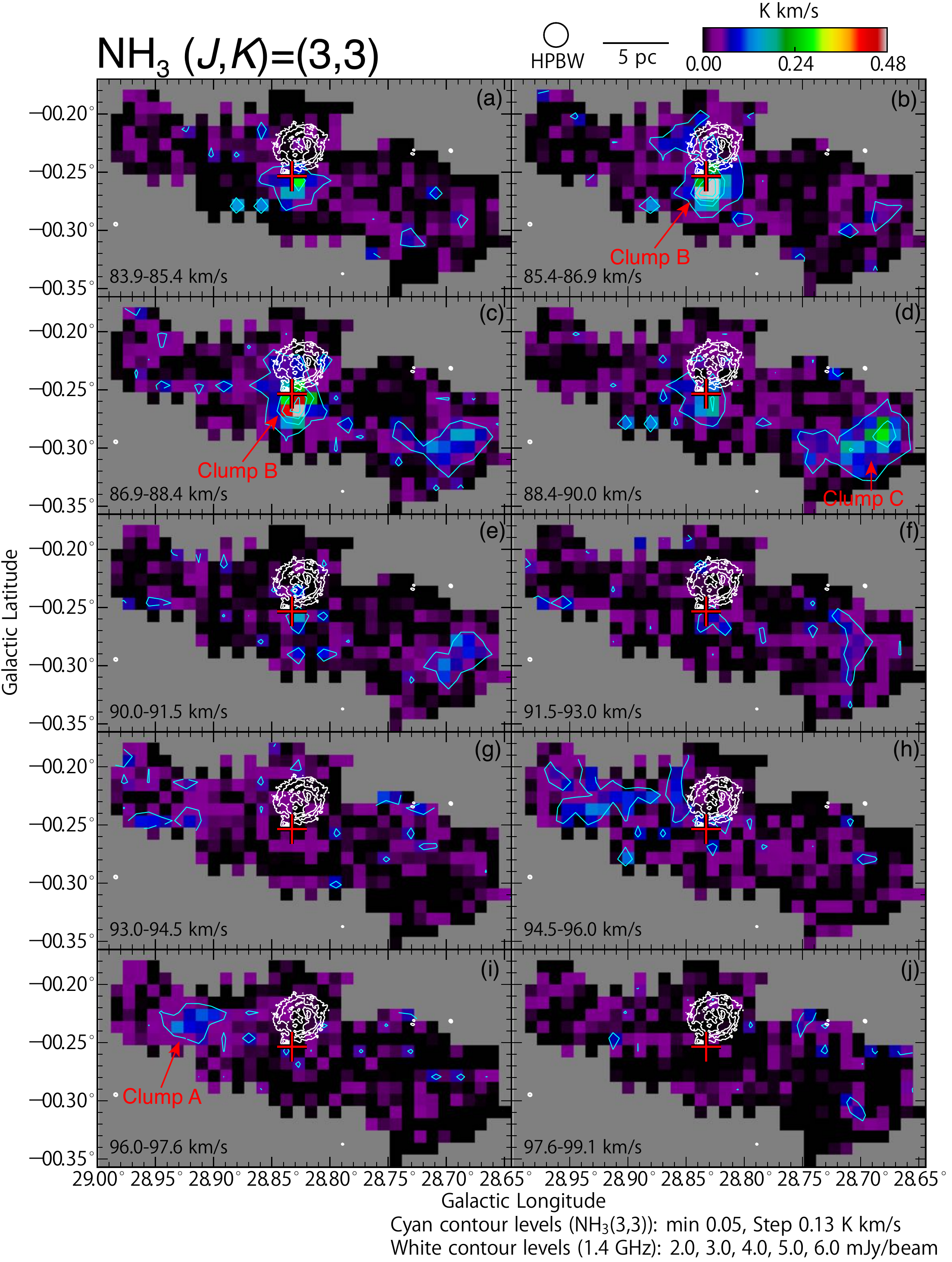}
\end{center}
\caption{Same as Figure \ref{ch}, but for NH$_3$ $(J,K)=(3,3)$.}
\label{ch3}
\end{figure*}

Figure \ref{ch}, \ref{ch2}, and \ref{ch3} show the \nh (1,1), \nh (2,2) ,and \nh (3,3) velocity channel map, respectively.
Clump A, B, and C has an intensity peak of all transitions at 96-97 \kms (panel i in  Figure \ref{ch}-\ref{ch3}), 85-88 \kms (panel b and c in Figure \ref{ch}-\ref{ch3}), and 88-90 \kms (panel d in Figure \ref{ch}-\ref{ch3}), respectively.
Clump A and B connect with the diffuse \nh (1,1) emission of 94-97 \kms (Figure \ref{ch}h and \ref{ch}i). 
To investigate the velocity field of each clump in more detail, we made the peak velocity and line width map of the \nh (1,1) main line obtained by a Gaussian fitting toward a profile at each pixel.

\clearpage

\begin{figure*}
\begin{center} 
\includegraphics[width=17cm]{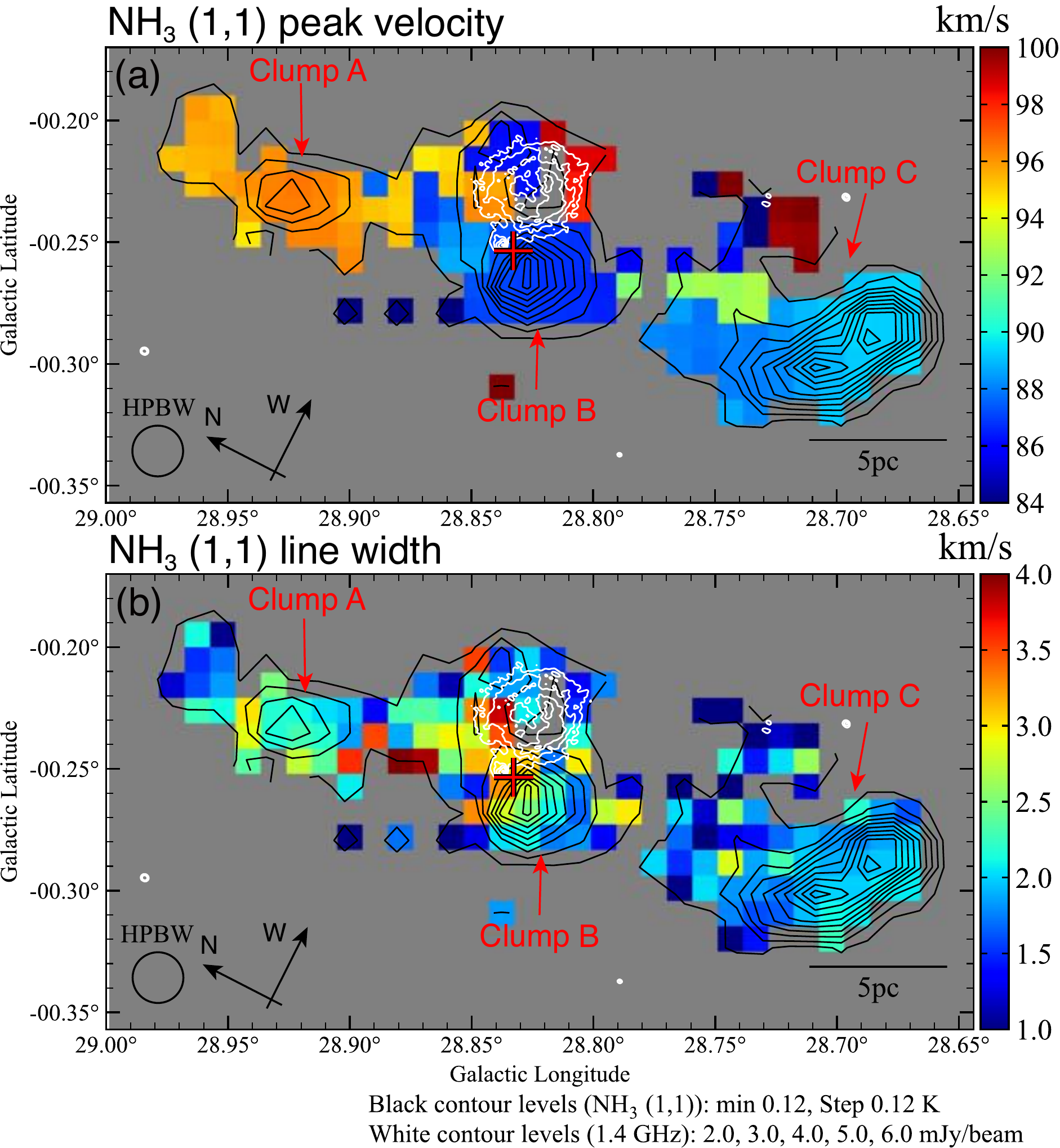}
\end{center}
\caption{{(a) Peak velocity and (b) velocity width maps of {the \nh $(J,K) = (1,1)$ line resulting from a Gaussian function fitted to the emission within pixels above {0.12 K}} ($> 4\sigma)$.
 The black contours show the \nh $(J,K) = (1,1)$ peak intensity. The lowest black contour level and intervals are  {0.12 K} ($\sim 4\sigma$). The red cross and white contours are the same as {in} Figure \ref{peakt}a. {The white contour levels of the VLA 1.4 GHz continuum image are 2.0, 3.0, 4.0, 5.0, and 6.0 mJy/beam.}}}
\label{peakdeltav}
\end{figure*}

Figure \ref{peakdeltav}(a) and (b) shows the peak velocity and line width map of the \nh (1,1) main line, respectively. 
Clump A, B, and C have different radial velocities at $\sim 96$, $\sim 87$, and $\sim 89$ \kms (Figure \ref{peakdeltav}a).
We also find the 100 \kms velocity component at $(l,b) \sim (\timeform{28.72D}, \timeform{-0.24D})$ around the northwestern side of Clump C.
The line width of Clump A is $\sim 2.3$ \kms.
Clump B has a large line width of $\sim 3.0$ \kms around the methanol maser source shown in the cross mark.
Finally, we point out that Clump C has a uniform line width of $\sim 2.0$ \kms.

\subsection{Physical properties of \nh clumps}

{Following procedures, we estimated the physical parameters of three \nh clumps from our \nh data  \citep{1983ARA&A..21..239H,1986A&A...157..207U, 1992ApJ...388..467M,2015PASP..127..266M}.
{First of all, we defined the clump radius ($r$) by
\begin{eqnarray}
r &= \sqrt{S \over \pi},
\label{eq:size}
\end{eqnarray}
where $S$ is the clump area detected above $3 \sigma$, assuming spherical shapes of the \nh (1,1) and \nh (2,2) emission.}
The {radius} of Clump A, B, and C is $\sim 2$ pc, $\sim 3$ pc, and $\sim 3$ pc, respectively.}

{Then, we estimated the optical depth ($\tau$), rotational temperature ($T_{\rm rot}$), and total column density ($N_{\rm TOT}$).}
The optical depth [$\tau(1,1,m)$] of the \nh (1,1) main line was derived from the intensity ratio of the main and satellite line above $2\sigma$ using the following equation.

\begin{eqnarray}
{{T_{\rm MB}}({\rm main}) \over {T_{\rm MB}}({\rm sate}) }={1-e^{-\tau({\rm 1,1,m})} \over 1-e^{-a\tau({\rm 1,1,m})}},
\label{eq:tau}
\end{eqnarray}
where {$T_{\rm MB}$ is the main-beam temperature}, $a$ is the theoretical intensity ratio, the main to the inner and outer-satellite lines are $a=0.278$ and $a=0.222$, respectively \citep{2002ApJ...577..757M}. 
{In this paper, we adopted the median value of four optical depths derived from the ratios of the main to four satellite lines. 
The mean optical depth in Clump A, B, and C are $1.4 \pm 0.6, 1.6 \pm 0.7,$ and $1.7 \pm 0.9$, respectively. The error shows the standard deviations of the mean value.}

\begin{figure*}
\begin{center} 
\includegraphics[width=17cm]{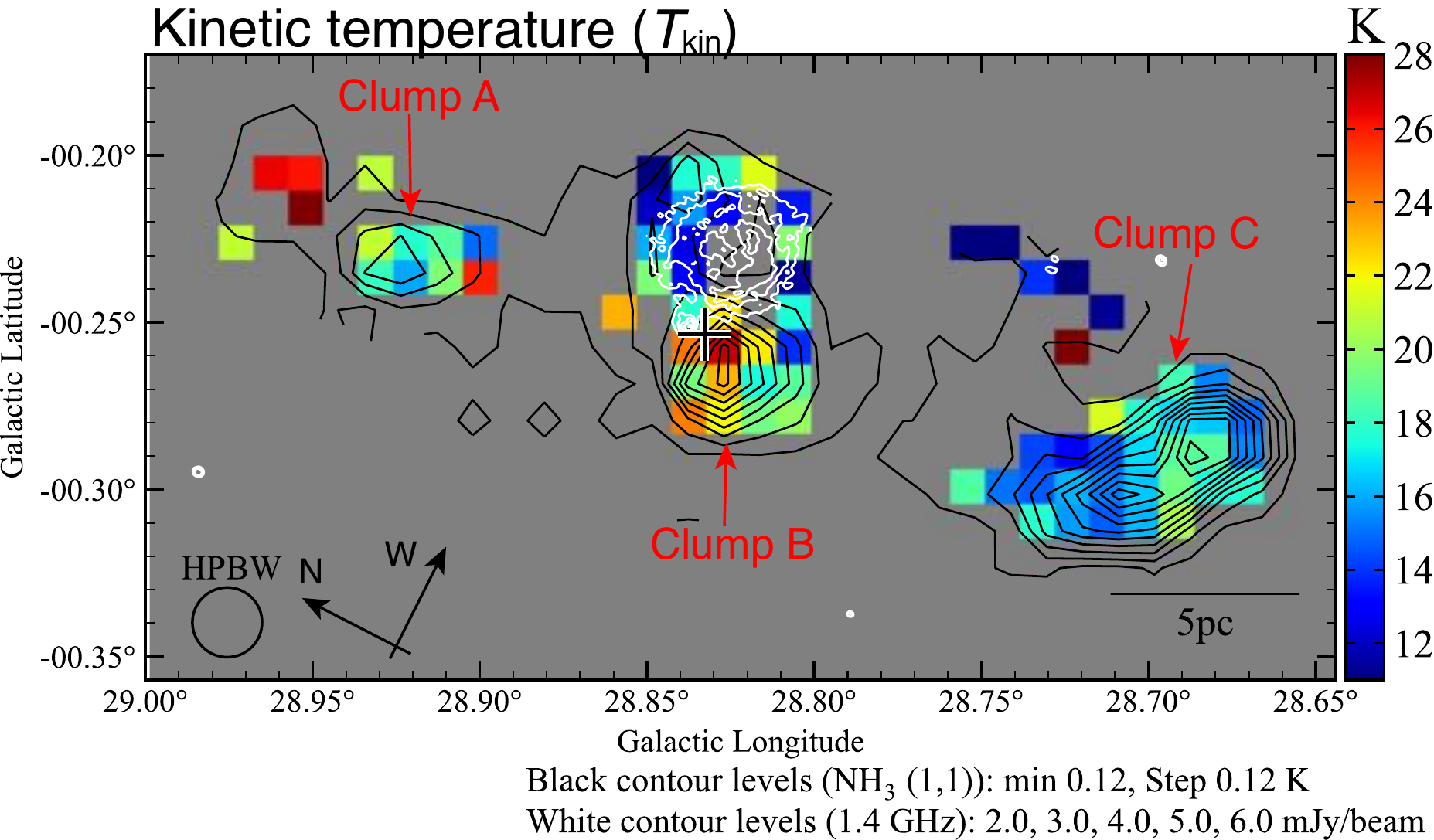}
\end{center}
\caption{{The kinetic temperature map obtained by the (2,2)/(1,1) integrated intensity ratio above the $3\sigma$ noise level. The black contours show the \nh $(J,K) = (1,1)$ peak intensity. The lowest black contour level and intervals are {0.12 K} ($\sim 4\sigma$). The black cross and white contours are the same as {in} Figure \ref{peakt}a. {The white contour levels of the VLA 1.4 GHz continuum image are 2.0, 3.0, 4.0, 5.0, and 6.0 mJy/beam.}}}
\label{tkin}
\end{figure*}

We estimated the rotational temperature from the \nh (2,2)/(1,1) intensity ratio of the main line detected above $3\sigma$. 
If we assume that the same line widths of \nh (1,1) and \nh (2,2) (e.g., \cite{2011MNRAS.418.1689U,2018A&A...609A.125W}), the rotational temperature is given by

\begin{eqnarray}
T_{\rm rot}(2,2:1,1) &= -41.1 \bigg/ \ln \left[\left({-0.282 \over \tau(1,1,m)}\right)\times \ln \left \{1-{{T_{\rm MB}}(2,2,m) \over {T_{\rm MB}}(1,1,m)}\times (1-e^{-\tau(1,1,m)} )\right\}\right]\  [{\rm K}].
\nonumber
\\
\label{eq:trot}
\end{eqnarray}
We converted the rotational temperature to kinetic temperature ($T_{\rm kin}$) by adopting the Monte Carlo method (Appendix B of \cite{2004A&A...416..191T}).
{The error of the kinetic temperature and rotational temperature was calculated by the error of the Gaussian fitting toward the peak intensity of \nh (1, 1), \nh (2, 2), and satellite lines.}

\begin{eqnarray}
T_{\rm kin} &= T_{\rm rot} \bigg/ \left\{1- {T_{\rm rot} \over 41.1} \ln \left[1+ 1.1 \exp \left(-{16 \over T_{\rm rot}} \right) \right]\right\}\  [{\rm K}]
\label{eq:tkin}
\end{eqnarray}
Figure \ref{tkin} presents the kinetic temperature distribution.
{The center of Clump A has $T_{\rm kin} = 17.7 \pm 0.1$ K, and the northern part at $(l,b)\sim (\timeform{28.96D}, \timeform{-0.206D})$ has a high temperature of $T_{\rm kin} = 26.4 \pm 0.3$ K.}
Clump B around the methanol maser source is enhanced of {$T_{\rm kin} = 27.0 \pm 0.6$ K}.
This peak value is consistent with $T_{\rm dust} \sim 27$ K obtained by the Herschel continuum image (see Figure 8a in \cite{2017ApJ...851..140D}). 
Clump C has uniform temperature distribution of $T_{\rm kin} \sim 14$-$18$ K.

We calculated the \nh column density from the \nh (1,1) and \nh (2,2) data points above $3\sigma$.
The \nh (1,1) column density assuming Local Thermal Equilibrium (LTE) is given by

\begin{eqnarray}
N(1,1) &= 2.78 \times 10^{13} \tau (1,1,m)\left({T_{\rm rot} \over [\rm K]}\right)\left({\Delta v \over [\rm km\ s^{-1}]}\right) [{\rm cm^{-2}}],
\label{eq:n1}
\end{eqnarray}
where $\Delta v$ is the line width of the main line.
Furthermore, the \nh column density \citep{1991ApJS...76..617T,2013tra..book.....W} summing up the total energy levels is given by 

\begin{eqnarray}
N_{\rm TOT}(\rm NH_3) &=& {N(J,K) \over g_J \cdot g_I \cdot g_K} Q_{\rm rot} \exp \left({E_u(J,K) \over kT_{\rm rot}}\right) \\
 &\sim& N(1,1) \Biggr[{1 \over 3}\exp \left({23.3 \over T_{\rm rot}}\right) + 1 + {5 \over 3}\exp \left({-41.1 \over T_{\rm rot}}\right) + {14 \over 3}\exp \left({-100.2 \over T_{\rm rot}}\right)\Biggr],
\label{eq:ntot}
\end{eqnarray}
where $Q_{\rm rot}$ is the partition function expressed as follows.

\begin{eqnarray}
Q_{\rm rot} = {\displaystyle \sum_{J}  \sum_{K} } g_J \cdot g_I \cdot g_K  \exp \left(-{E_u(J,K) \over kT_{\rm rot}}\right),
\label{eq:part}
\end{eqnarray}
where $ g_J$ is rotational degeneracy, $g_I$ is nuclear spin degeneracy, $g_K$ is K-degeneracy, $k$ is Boltzmann constant, and $E_u(J,K)$ is the upper energy level of inversion transition from the grand state, respectively.
{The uncertainties of total column densities were estimated by taking into account errors in the rotational temperature.}

\begin{figure*}
\begin{center} 
\includegraphics[width=17cm]{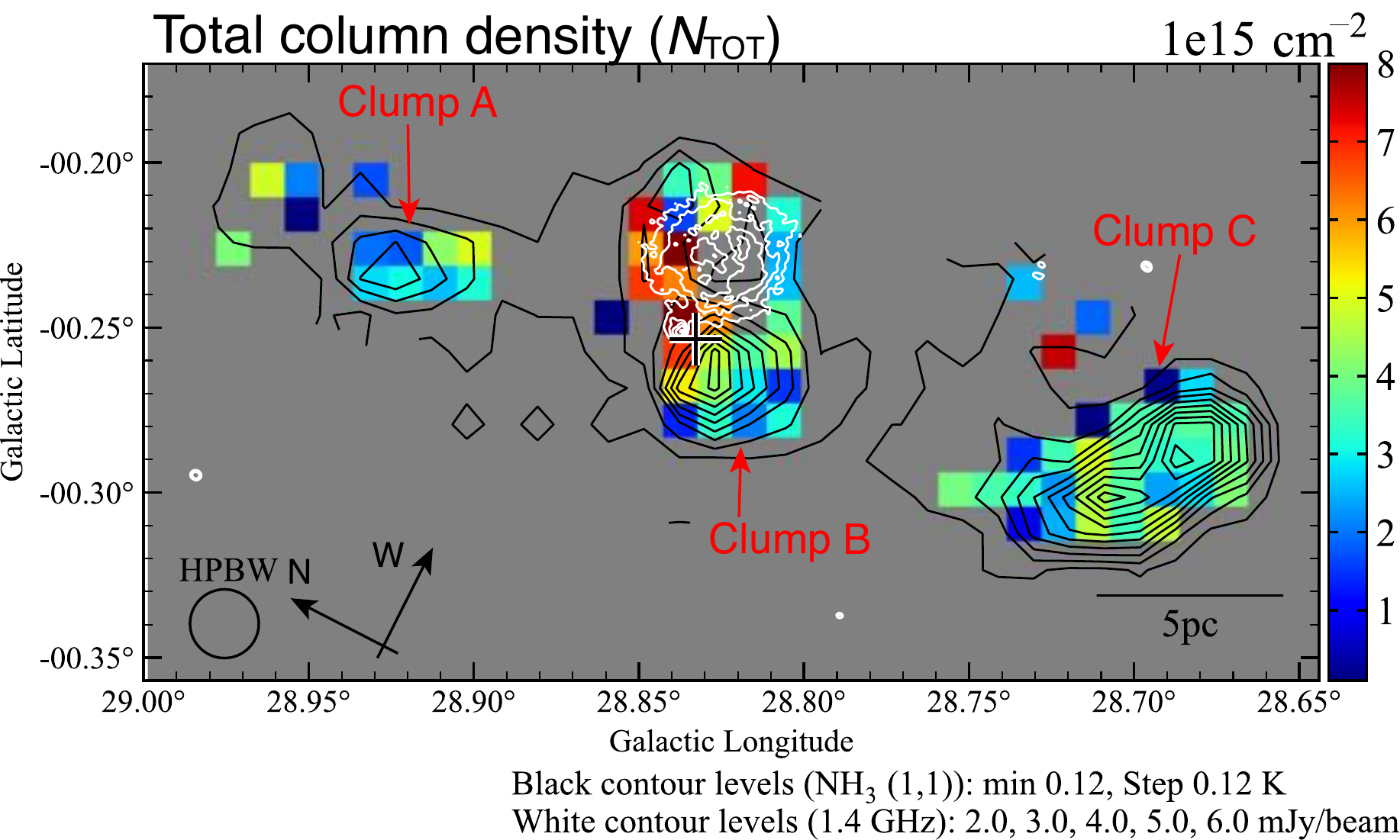}
\end{center}
\caption{{The \nh total column density map above the $3\sigma$ noise level. The contours show the \nh $(J,K) = (1,1)$ peak intensity. The lowest black contour level and intervals are {0.12 K} ($\sim 4\sigma$). The black cross and white contours are the same as {in} Figure \ref{peakt}a. {The white contour levels of the VLA 1.4 GHz continuum image are 2.0, 3.0, 4.0, 5.0, and 6.0 mJy/beam.}}}
\label{ntot}
\end{figure*}

Figure \ref{ntot} shows the \nh total column density map.
Clump B has a high column density of {$(9.1 \pm 0.4) \times 10^{15}$ cm$^{-2}$ at $(l,b) = (\timeform{28.837D}, \timeform{-0.247D})$, which coincides with the UC\HII region. In contrast, Clump A and C have a low of $ (1.8^{+0.09}_{-0.08}) \times 10^{15}$ cm$^{-2}$ and $\sim (3.7 \pm 0.3) \times 10^{15}$ cm$^{-2}$, respectively.
These values correspond to $\sim 3.6\times 10^{23}$ cm$^{-2}$, $\sim 7.2\times 10^{22}$ cm$^{-2}$, and $\sim 1.5\times 10^{23}$ cm$^{-2}$ of the H$_2$ column density, respectively. }
We used the $N({\rm NH_3})$ to $N({\rm H_2}$) conversion factor of $X({\rm NH_3})=2.5 \times 10^{-8}$ derived from the massive star-forming regions in the Galactic plane \citep{2015MNRAS.452.4029U}. 
The peak H$_2$ column density around the methanol maser source is factor 4 larger than $\sim 9 \times 10^{22}$ cm$^{-2}$ obtained by the Herschel dust continuum image \citep{2017ApJ...851..140D}.

\begin{table*}
{
\tbl{Physical parameters in each \nh clump}{
\begin{tabular}{cccccccccc}
\hline
\multicolumn{1}{c}{Name}& {Radius} & ${\tau (1,1,m)}$ & ${T_{\rm rot}}$ & ${T_{\rm kin}}$ &  ${N_{\rm tot}}$(NH$_3$)&  ${N({\rm H_2})}$  & $M_{\rm LTE}$ 
\\
& [pc]  &  & [K] & [K] & [cm$^{-2}$] & [cm$^{-2}$]  & [$M_{\odot}$] 
 \\
(1) & (2) & (3) &(4)& (5) & (6) & (7)&(8)
\\
\hline
Clump A & 1.6 & $1.4 \pm 0.6$ & $16.5 \pm 2.0$ & $19.2 \pm 3.0$ & ($2.9 \pm 1.0)\times 10^{15}$  & $(1.2\pm 0.4) \times 10^{23}$ & $2.0 \times 10^4$
\\
Clump B & 2.8 & $1.6 \pm 0.7$ & $15.4 \pm 3.9$ & $17.9 \pm 5.2$ & $(4.3 \pm 2.2) \times 10^{15}$  & $ (1.7 \pm 0.9) \times 10^{23}$  & $9.5\times 10^4$ 
\\
Clump C & 2.8 & $1.5\pm 0.6$ & $14.6 \pm 1.5$ & $16.5 \pm 2.1$ & $\left(3.2 \pm 1.2 \right) \times 10^{15}$ & $ \left(1.3 \pm 0.5 \right) \times 10^{23}$  & $7.1\times 10^4$
\\
\hline
\end{tabular}}\label{physic_param}
\begin{tabnote}
Columns: (1) Clump name (2) The clump {radius} defined above $3 \sigma$ of the \nh (1,1) and \nh (2,2) emission.  (3) The mean value of optical depth. (4) The mean value of rotational temperature. (5) The mean value of the kinetic temperature. (6) The mean value of the total \nh column density. (7) The mean value of H$_2$ column density. (8) The total LTE mass of a clump. The errors are derived by the standard deviation of the mean value in each clump.
\end{tabnote}
}
\end{table*}

The LTE mass is estimated by 
\begin{eqnarray}
M_{\rm LTE} &= \mu_{\rm H_2} m_{\rm H} D^2 \sum_{i} \Omega\ N_i(\rm{H_2}),
\label{eq:mass}
\end{eqnarray}
where $ \mu_{\rm H_2}\sim 2.8 $ is the mean molecular weight {per hydrogen molecule} (Appendix A.1 in \cite{2008A&A...487..993K}), $m_{\rm H} = 1.67 \times 10^{-24}$g is the proton mass, $D=5.07$ kpc is the distance to N49, $N_i(\rm{H_2})$ is the $\rm{H_2}$ column density at the {$i$th pixel}, and $\Omega$ is the solid angle of each grid.
The LTE mass of Clump A, Clump B, and Clump C is {$2.0 \times 10^4\ M_{\odot}, 9.5 \times 10^4\ M_{\odot}$, and $7.1 \times 10^4\ M_{\odot}$, respectively. }
\citet{2017ApJ...851..140D} identified the dust clumps using the Herschel far-infrared continuum data (see Figure 9 and Table 1 in \cite{2017ApJ...851..140D}). 
Our identified three \nh clumps of A, B, and C correspond to their dust {clumps} of \#15 ($M_{\rm clump} = 5.5 \times 10^3\ M_{\odot}$), \#12-\#13 ($M_{\rm clump} = 2.0 \times 10^4\ M_{\odot}$), and \#7-\#8-\#9 ($M_{\rm clump} = 1.5 \times 10^4\ M_{\odot}$), respectively. 
{We point out that the mass of clump \#12-\#13 and clump \#7-\#8-\#9 is the sum of the individual Herschel clumps in \citet{2017ApJ...851..140D}.}
The LTE mass calculated from \nh data is factor 4-5 larger than the dust clump mass estimated by \citet{2017ApJ...851..140D}.
{Table \ref{physic_param} presents the physical parameters of each \nh clump. The errors were estimated by the standard deviation of the mean value in each clump.}

\section{Discussion}
\subsection{Distributions of cold dust condensations and star formation in \nh clumps}
\begin{figure*}
\begin{center} 
\includegraphics[width=18cm]{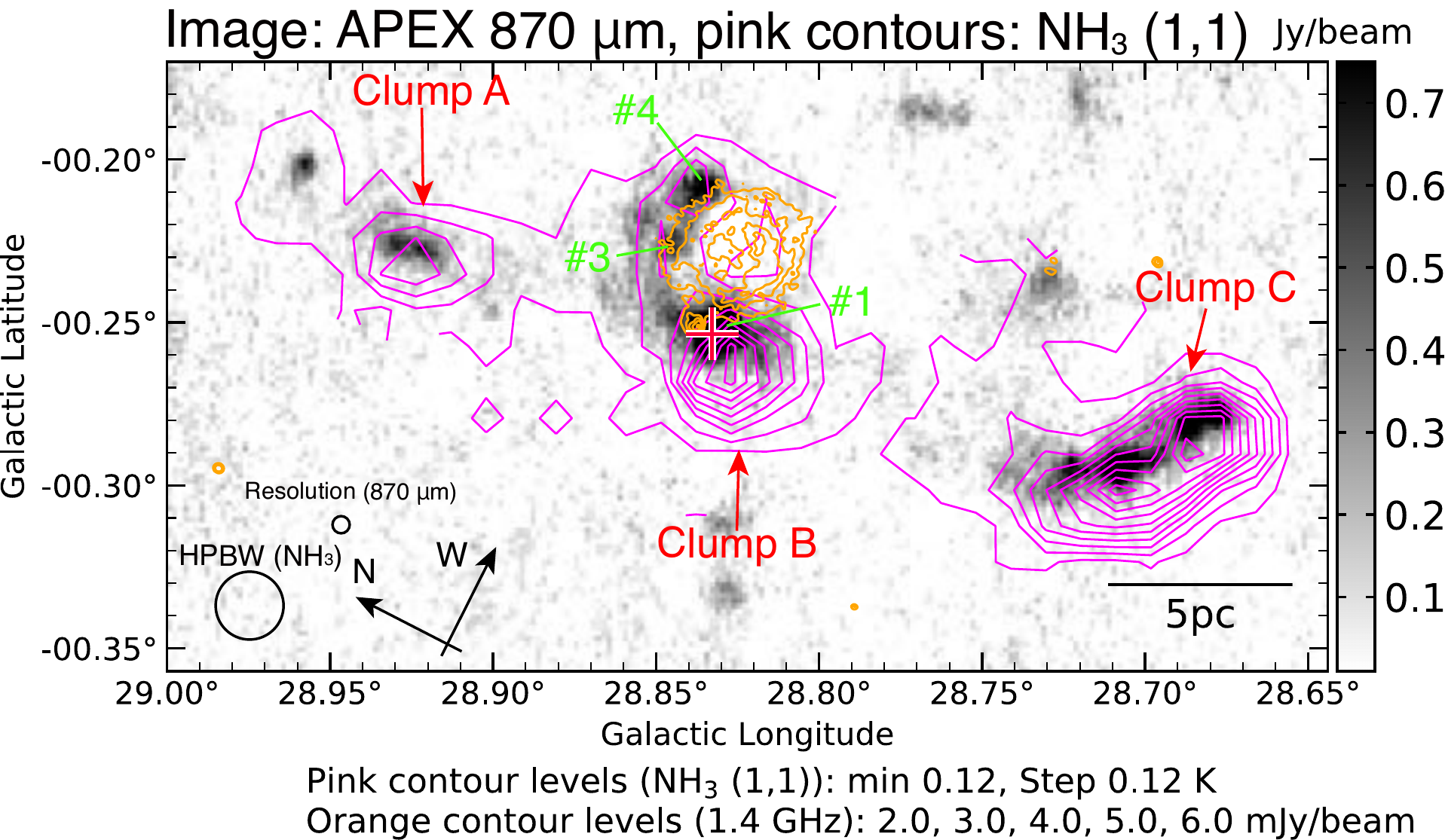}
\end{center}
\caption{{{The \nh $(J, K) = (1, 1)$ peak intensity shown as pink contours} superposed on the APEX 870 $\mu$m continuum image. The lowest pink contour level and intervals are {0.12 K} ($\sim 4\sigma$). The red cross is the same as {in} Figure \ref{peakt}a. {The orange contour levels of the VLA 1.4 GHz continuum image are 2.0, 3.0, 4.0, 5.0, and 6.0 mJy/beam.} \#1, \#3, and \#4 indicate the dust condensations identified by {\citet{2010A&A...523A...6D}.}}}
\label{870}
\end{figure*}

Figure \ref{870} shows the \nh distributions presented by pink contours superposed on the 870 $\mu$m dust continuum image. 
{We point out Clump A, B, and C corresponding to the dust condensations of AGAL028.922-00.227\footnote{\url{https://atlasgal.mpifr-bonn.mpg.de/cgi-bin/ATLASGAL_SEARCH_RESULTS.cgi?text_field_1=AGAL028.922-00.227}}, AGAL028.831-00.252\footnote{\url{https://atlasgal.mpifr-bonn.mpg.de/cgi-bin/ATLASGAL_SEARCH_RESULTS.cgi?text_field_1=AGAL028.831-00.252}}, and AGAL028.707-00.294\footnote{\url{https://atlasgal.mpifr-bonn.mpg.de/cgi-bin/ATLASGAL_SEARCH_RESULTS.cgi?text_field_1=AGAL028.707-00.294}}, respectively. 
These dust clumps were identified by the ATLASGAL survey (\cite{2014A&A...565A..75C,2014A&A...568A..41U}). The dust peak also exists at $(l,b) \sim (\timeform{28.96D}, \timeform{-0.20D})$, which is the northern side of Clump A. 
The \nh peaks of Clump B coincide with three dust condensations of \#1, \#3, and \#4 along with the edge of the N49 bubble (see also Figure 18 in \cite{2010A&A...523A...6D}).}

\begin{figure*}
\begin{center} 
\includegraphics[width=18cm]{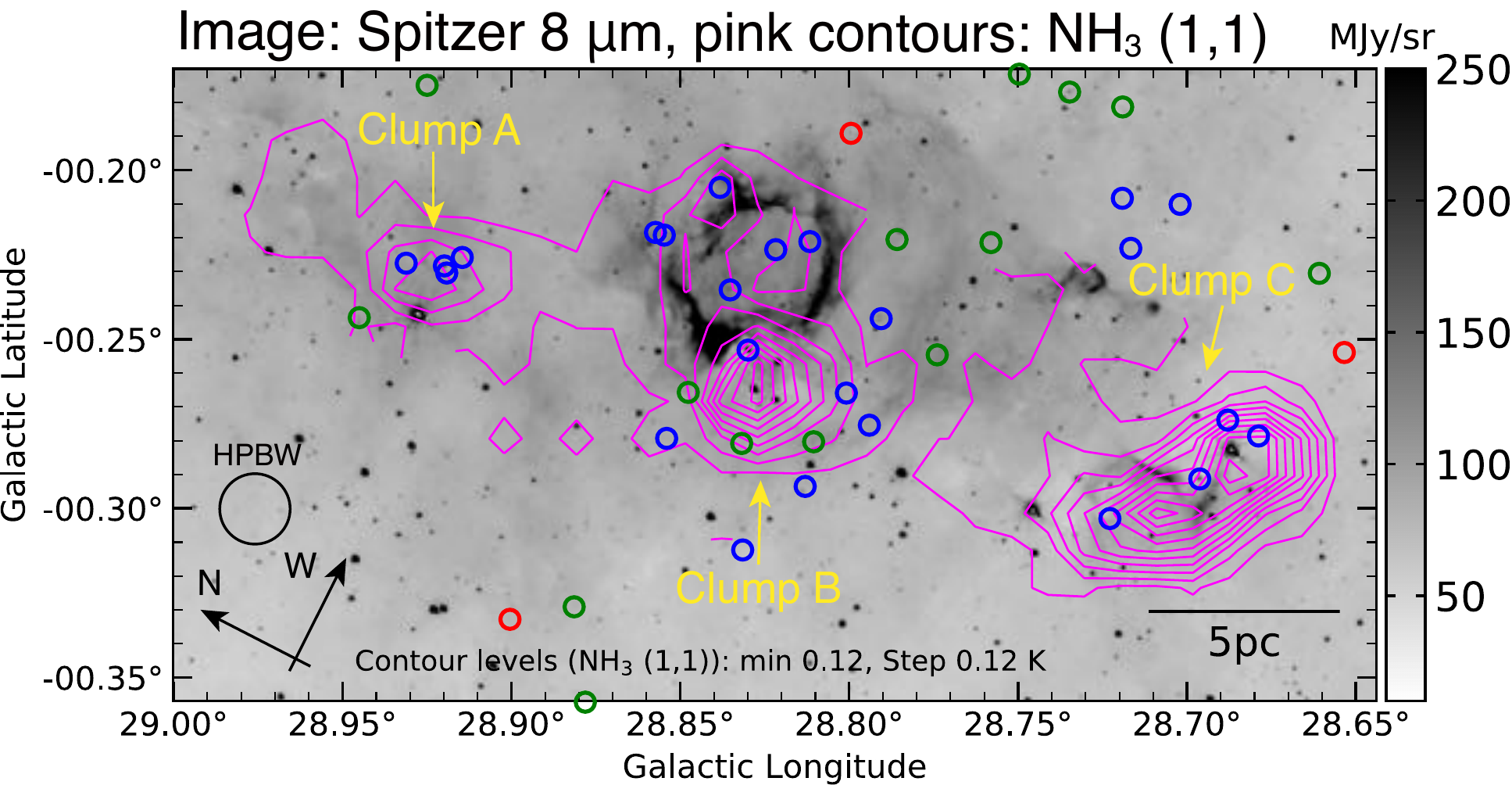}
\end{center}
\caption{{YSO distributions superposed on the Spitzer 8 $\mu$m continuum image. Pink contours indicate the \nh $(J,K) = (1,1)$ peak intensity. The lowest pink contour level and intervals are {0.12 K} ($\sim 4\sigma$). 
Blue, green, {red} circles show the Stage I, Stage II, and Stage III YSOs identified by \citet{2012AJ....144..173D} based on the classification model of \citet{2006ApJS..167..256R}.}}
\label{yso}
\end{figure*}

Figure \ref{yso} presents the distributions of YSOs identified by \citet{2012AJ....144..173D}.
The image and pink contours show the Spitzer 8 $\mu$m and peak intensity of \nh (1,1), respectively.
\citet{2012AJ....144..173D} classified YSOs to ``Stage I, Stage II, and Stage III" with their evolutional stage defined by \citet{2006ApJS..167..256R}.
Clump A and C have Stage I YSOs shown in blue circles. 
Clump B has Stage I YSOs at the edge and inside the bubble. 
Stage II YSOs exist on the eastern side of Clump B.
We suggest that the low mass star formation proceeds at each \nh clump in the molecular filament because YSOs exist at each dense clump.
{Clump A and C exists at $\sim 5$ pc and $\sim 10$ pc away from the edge of the bubble traced by the 1.4 GHz radio continuum image.  Therefore, we propose that these clumps might not be affected by the external shock compression or ionization from the N49 bubble, but the sites of spontaneous star formation.}
{Clump C has low kinetic temperatures of 14-18 K (Figure \ref{tkin}) and the uniform line width of $\sim 2.0$ \kms(Figure \ref{peakdeltav}b), while it includes \nh dense gas, cold dust condensation, and Stage I YSOs (Figure \ref{870}). 
We suppose that they show signatures of the early stage of star formation in Clump C.}
{Clump A and C are not detected in the radio continuum emission even though dust condensations and YSOs. Therefore, these clumps might be the more initial stage or small scale of star formation than Clump B.}

\subsection{Impact of the N49 bubble to the \nh clump}
{The \nh (1,1) main line has a large line width of $3.8 \pm 0.3$ \kms at $(l,b) = (\timeform{28.837D},\timeform{-0.226D})$, which corresponds to the northern interface of Clump B and ionized gas (Figure \ref{peakdeltav}b). 
{This result suggests that the turbulent motion is enhanced locally by the interaction of gas within the bubble and the \nh dense clump.}
The distributions of kinetic temperature have a local high $T_{\rm kin}= 27.0 \pm 0.6$ K at the 6.7 GHz methanol maser source associated with MYSOs \citep{2009ApJ...702.1615C}, while it has a low $T_{\rm kin} < 18$ K at the northern and western edge of the bubble (Figure \ref{tkin}).
We propose that this shows the feedback from the young ($< 10^5$ yr) embedded MYSOs are heating the dense clump locally.
Our previous studies support this result (e.g., Monkey Head nebula{:} \cite{2013ApJ...762...17C}, AFGL-333{:} \cite{2017PASJ...69...16N}, W33{:} \cite{2021arXiv211113481M}, and Sh 2-255{:} \cite{2022S255}). These papers concluded that embedded cluster and the compact \HII region with their age of $< 10^5$ yr can {affect the local heating of the dense \nh clump.}}

\subsection{Two molecular filaments and triggered star formation}
\begin{figure*}
\begin{center} 
\includegraphics[width=12cm]{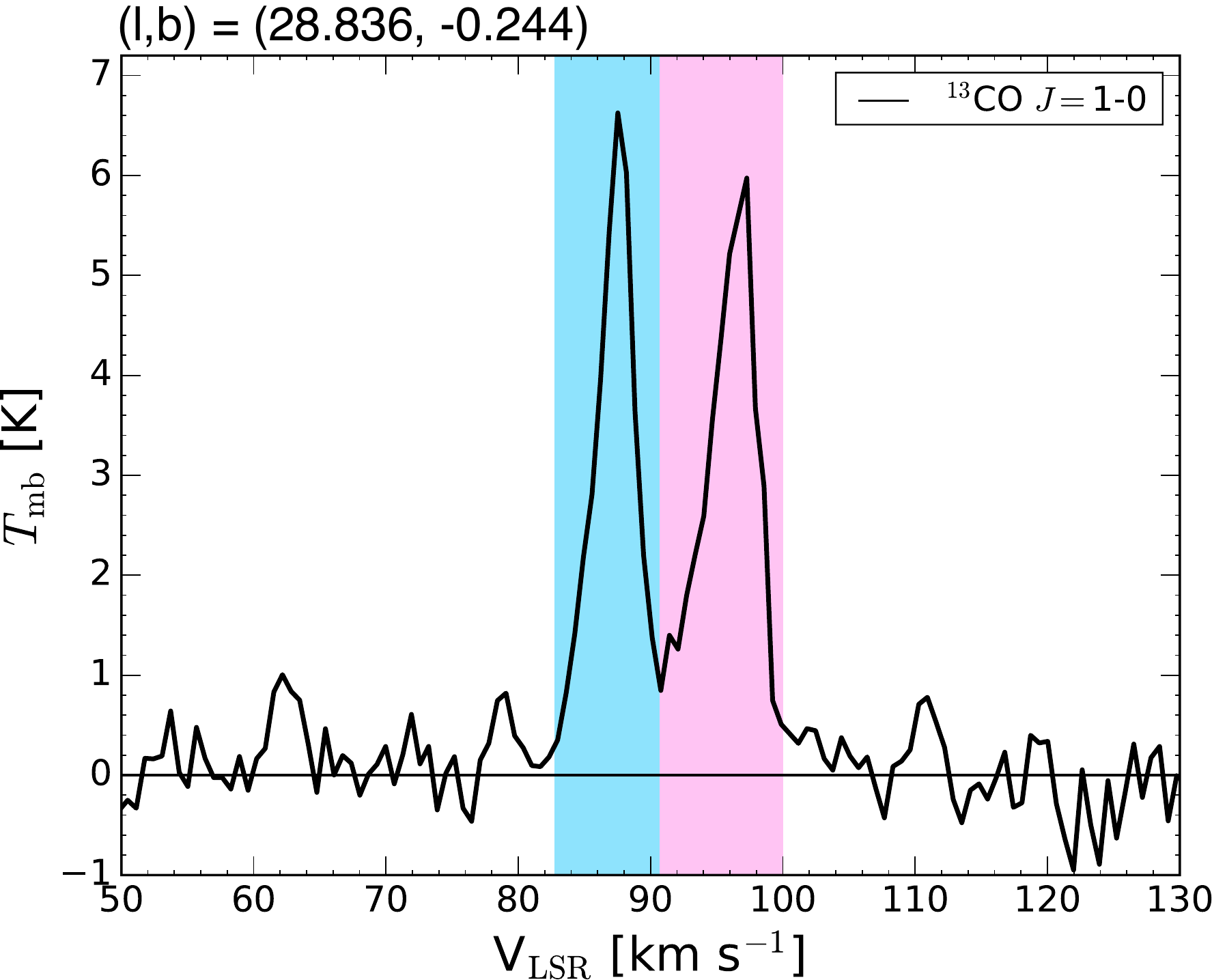}
\end{center}
\caption{{The FUGIN $^{13}$CO spectra at $(l,b)=(\timeform{28.836D}, \timeform{-0.244D})$. Blue and pink area show the integrated velocity range of the two molecular filaments in Figure \ref{COfilament}(a) and \ref{COfilament}(b), respectively. }}
\label{COspec}
\end{figure*}

\begin{figure*}[ht]
\begin{center} 
\includegraphics[width=18cm]{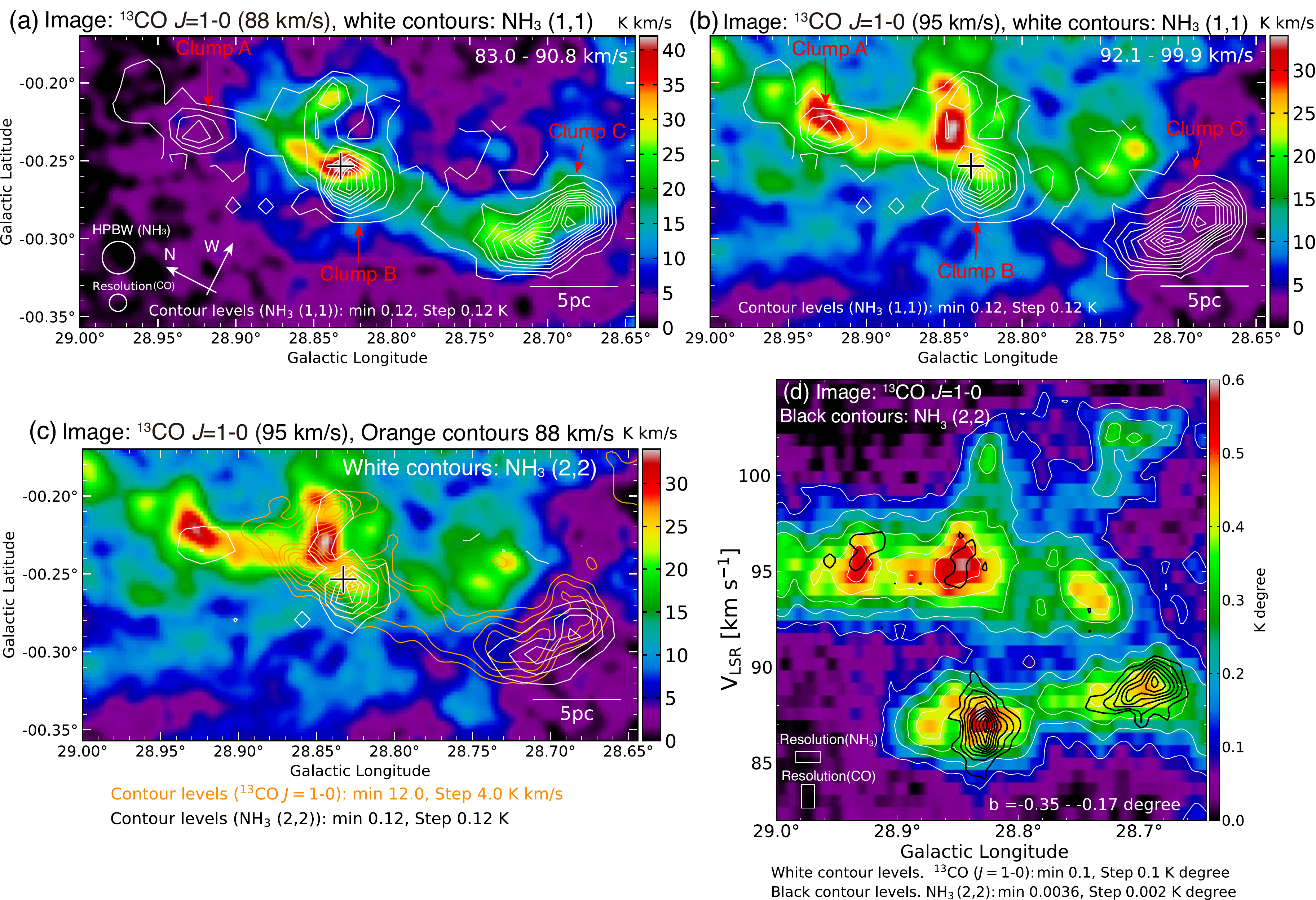}
\end{center}
\caption{{The FUGIN $^{13}$CO integrated intensity map of two molecular filaments identified by \citet{2017ApJ...851..140D}. The integrated velocity range of (a) 83.0-90.8 \kms and (b) 92.1-99.9 \kms, respectively. White contours indicate the \nh $(J,K) = (1,1)$ peak intensity. The lowest white contour level and intervals are {0.12 K} ($\sim 4\sigma$). (c) The integrated intensities of the 88 \kms component (orange contours) and 95 \kms components (background image). White contours indicate the \nh $(J,K) = (2,2)$ peak intensity. 
The black cross is the same as {in} Figure \ref{peakt}a. (d) {The longitude-velocity diagram of $^{13}$CO $J=$1-0 shown as white contours.}
The lowest white contour level and intervals are 0.1 K degree. The black contours present the \nh (2,2) intensity of the same integrated latitude range. The lowest black contour level and intervals are {0.0036 K} degree and 0.002 K degree, respectively.}} 
\label{COfilament}
\end{figure*}

Figure \ref{COspec} shows the FUGIN $^{13}$CO $J=$1-0 spectra at the edge of the bubble.
We can find the two velocity components of $\sim 88$ \kms and $\sim 95$ \kms.
\citet{2017ApJ...851..140D} reported the molecular filaments composed of these two velocity components from the $^{13}$CO $J=$1-0 Galactic Ring Survey data \citep{2006ApJS..163..145J}.
Figure \ref{COfilament}(a) and (b) present \nh peak intensity distributions superposed on the molecular filaments of $\sim 88$ \kms and $\sim 95$ \kms components, respectively.
The \nh peak of Clump B and C corresponds to the CO peak of the $\sim 88$ \kms component.
Clump A coincides with the CO peak of the $\sim 95$ \kms component.
Figure \ref{COfilament}(c) demonstrates \nh (2,2) peak intensity overlaid on the two velocity components shown in the image of 95 \kms and orange contours of 88 \kms.
Clump B exists at the overlapped region of the two velocity components.
This result supports the filament-filament collision-induced dense gas and massive star formation at the edge of the bubble proposed by \citet{2017ApJ...851..140D}.
Figure \ref{COfilament} (d) shows the longitude-velocity diagram. Black contours indicate the \nh (2,2) distribution. 
The 88 \kms and 95 \kms component connect with the bridge feature on the velocity space reported by  \citet{2017ApJ...851..140D}. This is evidence of the collisional interaction of two clouds from the numerical simulation by \citet{2015MNRAS.450...10H} based on the model of \citet{2014ApJ...792...63T}.
The \nh (2,2) peaks coincide with the $^{13}$CO dense points of the 88 \kms and 95 \kms component.
The \nh (2,2) intensity is more enhanced at {88 \kms} than the 95 \kms component.
We suggest that the dense gas produced by the shocked compression of filament collision is likely to exist at the {88 \kms\ component}. 
Indeed, the numerical simulation showed that the shock compression by cloud-cloud collision induces dense gas formation traced by \nh \citep{2021MNRAS.506..775P}. 
Therefore, our \nh results support that the current massive star and dense gas formation at the edge of the bubble was triggered by the collision event suggested by \citet{2017ApJ...851..140D}.

\section{Summary}
The conclusions of our paper are summarized as follows:
\begin{enumerate}
\item We have performed the \nh $(J,K)=(1,1),(2,2),$ and $(3,3)$ survey toward the Galactic infrared bubble N49 (G28.83-0.25) as a part of the KAGONMA project led by the Kagoshima University. 
\item We found three \nh clumps (A, B, and C) along the molecular filament with the radial velocities of $\sim$ 96, 87, and 89 \kms, respectively. 
\item The kinetic temperature derived from the \nh (2,2)/\nh (1,1) shows {$T_{\rm kin} = 27.0 \pm 0.6$ K} enhanced at Clump B in the eastern edge of the bubble, where position coincides with MYSOs associated with the 6.7 GHz methanol maser source. This result shows the dense clump is locally heated by the stellar feedback from the embedded MYSOs. 
\item The \nh Clump B also exists at the 88 \kms and 95 \kms molecular filament intersection. Therefore, we suggest that the filament-filament interaction scenario argued by \citet{2017ApJ...851..140D} can explain the \nh dense gas formation in Clump B.
{\item The \nh Clump A and C at the northern and southern sides of the molecular filament exist $\sim 5$ - $10$ pc apart from the edge of the bubble traced by the 1.4 GHz radio continuum image. This result suggests that these clumps are likely to occur in spontaneous star formation, not to be affected by the ionization feedback from the N49 bubble.}
\end{enumerate}

\section*{Acknowledgements}
{We are grateful to the anonymous referee for carefully reading our manuscript and giving us thoughtful suggestions, which greatly improved of the paper.}
The Nobeyama 45-m radio telescope is operated by Nobeyama Radio Observatory, a branch of the National Astronomical Observatory of Japan. The authors wish to thank all staff of the Nobeyama radio observatory for their assistance in our backup observations.
The NH$_3$ backup observations are promoted on a lot of contributions of Kagoshima university, so the authors would like to thank all them, Dr. Tatsuya Kamezaki, Dr. Gabor Orosz, Dr. Mitsuhiro Matsuo, Mr. Hideo Hamabata, Mr. Tatsuya Baba, Mr. Ikko Hoshihara, Mr. Kohei Mizukubo, Mr. Tatsuya Sera, and Mr. Masahiro Uesugi. 
I appreciated the curator members of the Astronomy section of Nagoya city science museum supporting my research works.
The data of the FUGIN project was retrieved from the JVO portal (\url{http://jvo.nao.ac.jp/portal/}) operated by ADC/NAOJ.  The data analysis was carried out at the Astronomy Data Center of the National Astronomical Observatory of Japan. The ATLASGAL project is a collaboration between the Max-Planck-Gesellschaft, the European Southern Observatory (ESO) and the Universidad de Chile. It includes projects E-181.C-0885, E-078.F-9040(A), M-079.C-9501(A), M-081.C-9501(A) plus Chilean data.
This work is based in part on observations made with the Spitzer Space Telescope, which was operated by the Jet Propulsion Laboratory, California Institute of Technology under a contract with NASA.



\begin{thebibliography}{}
\bibitem[Astropy Collaboration et al.(2013)]{2013A&A...558A..33A} Astropy Collaboration, Robitaille, T.~P., Tollerud, E.~J., et al.\ 2013, \aap, 558, A33
\bibitem[Astropy Collaboration et al.(2018)]{2018AJ....156..123A} Astropy Collaboration, Price-Whelan, A.~M., Sip{\H{o}}cz, B.~M., et al.\ 2018, \aj, 156, 123
\bibitem[Baug et al.(2016)]{2016ApJ...833...85B} Baug, T., Dewangan, L.~K., Ojha, D.~K., et al.\ 2016, \apj, 833, 85. doi:10.3847/1538-4357/833/1/85
\bibitem[Benjamin et al.(2003)]{2003PASP..115..953B} Benjamin, R.~A., Churchwell, E., Babler, B.~L., et al.\ 2003, \pasp, 115, 953. doi:10.1086/376696
\bibitem[Burns et al.(2019)]{2019PASJ...71...91B} Burns, R.~A., Handa, T., Omodaka, T., et al.\ 2019, \pasj, 71, 91. doi:10.1093/pasj/psz074
\bibitem[Carey et al.(2009)]{2009PASP..121...76C} Carey, S.~J., Noriega-Crespo, A., Mizuno, D.~R., et al.\ 2009, \pasp, 121, 76. doi:10.1086/596581
\bibitem[Castor et al.(1975)]{1975ApJ...200L.107C} Castor, J., McCray, R., \& Weaver, R.\ 1975, \apjl, 200, L107. doi:10.1086/181908
\bibitem[Cheung et al.(1968)]{1968PhRvL..21.1701C} {Cheung, A.~C., Rank, D.~M., Townes, C.~H., et al.\ 1968, \prl, 21, 1701. doi:10.1103/PhysRevLett.21.1701}
\bibitem[Cheung et al.(1969)]{1969ApJ...157L..13C} {Cheung, A.~C., Rank, D.~M., Townes, C.~H., et al.\ 1969, \apjl, 157, L13. doi:10.1086/180374}
\bibitem[Chibueze et al.(2013)]{2013ApJ...762...17C} Chibueze, J.~O., Imura, K., Omodaka, T., et al.\ 2013, \apj, 762, 17. doi:10.1088/0004-637X/762/1/17
\bibitem[Churchwell et al.(2006)]{2006ApJ...649..759C} Churchwell, E., Povich, M.~S., Allen, D., et al.\ 2006, \apj, 649, 759. doi:10.1086/507015
\bibitem[Churchwell et al.(2007)]{2007ApJ...670..428C} Churchwell, E., Watson, D.~F., Povich, M.~S., et al.\ 2007, \apj, 670, 428. doi:10.1086/521646
\bibitem[Churchwell et al.(2009)]{2009PASP..121..213C} Churchwell, E., Babler, B.~L., Meade, M.~R., et al.\ 2009, \pasp, 121, 213. doi:10.1086/597811
\bibitem[Csengeri et al.(2014)]{2014A&A...565A..75C} Csengeri, T., Urquhart, J.~S., Schuller, F., et al.\ 2014, \aap, 565, A75. doi:10.1051/0004-6361/201322434
\bibitem[Cyganowski et al.(2008)]{2008AJ....136.2391C} Cyganowski, C.~J., Whitney, B.~A., Holden, E., et al.\ 2008, \aj, 136, 2391. doi:10.1088/0004-6256/136/6/2391
\bibitem[Cyganowski et al.(2009)]{2009ApJ...702.1615C} Cyganowski, C.~J., Brogan, C.~L., Hunter, T.~R., et al.\ 2009, \apj, 702, 1615. doi:10.1088/0004-637X/702/2/1615
\bibitem[Deharveng et al.(2005)]{2005A&A...433..565D} Deharveng, L., Zavagno, A., \& Caplan, J.\ 2005, \aap, 433, 565. doi:10.1051/0004-6361:20041946
\bibitem[Deharveng et al.(2010)]{2010A&A...523A...6D} Deharveng, L., Schuller, F., Anderson, L.~D., et al.\ 2010, \aap, 523, A6. doi:10.1051/0004-6361/201014422
\bibitem[Dewangan et al.(2017)]{2017ApJ...851..140D} Dewangan, L.~K., Ojha, D.~K., \& Zinchenko, I.\ 2017, \apj, 851, 140. doi:10.3847/1538-4357/aa9be2
\bibitem[Dirienzo et al.(2012)]{2012AJ....144..173D} Dirienzo, W.~J., Indebetouw, R., Brogan, C., et al.\ 2012, \aj, 144, 173. doi:10.1088/0004-6256/144/6/173
\bibitem[Draine \& Li(2007)]{2007ApJ...657..810D} Draine, B.~T. \& Li, A.\ 2007, \apj, 657, 810. doi:10.1086/511055
\bibitem[Draine(2003)]{2003ARA&A..41..241D} Draine, B.~T.\ 2003, \araa, 41, 241. doi:10.1146/annurev.astro.41.011802.094840
\bibitem[Dale et al.(2007)]{2007MNRAS.375.1291D} Dale, J.~E., Bonnell, I.~A., \& Whitworth, A.~P.\ 2007, \mnras, 375, 1291. doi:10.1111/j.1365-2966.2006.11368.x
\bibitem[Elmegreen \& Lada(1977)]{1977ApJ...214..725E} Elmegreen, B.~G. \& Lada, C.~J.\ 1977, \apj, 214, 725. doi:10.1086/155302
\bibitem[Elmegreen(1998)]{1998ASPC..148..150E} Elmegreen, B.~G.\ 1998, Origins, 148, 150
\bibitem[Enokiya \& Fukui(2022)]{2022ApJ...931..155E} {Enokiya, R. \& Fukui, Y.\ 2022, \apj, 931, 155. doi:10.3847/1538-4357/ac674f}
\bibitem[Everett \& Churchwell(2010)]{2010ApJ...713..592E} Everett, J.~E. \& Churchwell, E.\ 2010, \apj, 713, 592. doi:10.1088/0004-637X/713/1/592
\bibitem[Fukui et al.(2018)]{2018PASJ...70S..46F} Fukui, Y., Ohama, A., Kohno, M., et al.\ 2018, \pasj, 70, S46. doi:10.1093/pasj/psy005
\bibitem[Fukui et al.(2021)]{2021PASJ...73S...1F} Fukui, Y., Habe, A., Inoue, T., et al.\ 2021, \pasj, 73, S1. doi:10.1093/pasj/psaa103

\bibitem[Fujita et al.(2019)]{2019ApJ...872...49F} Fujita, S., Torii, K., Tachihara, K., et al.\ 2019, \apj, 872, 49. doi:10.3847/1538-4357/aafac5
\bibitem[Fujita et al.(2021)]{2021PASJ...73S.172F} Fujita, S., Torii, K., Kuno, N., et al.\ 2021, \pasj, 73, S172. doi:10.1093/pasj/psz028
\bibitem[Habe \& Ohta(1992)]{1992PASJ...44..203H} Habe, A. \& Ohta, K.\ 1992, \pasj, 44, 203
\bibitem[Hanaoka et al.(2019)]{2019PASJ...71....6H} Hanaoka, M., Kaneda, H., Suzuki, T., et al.\ 2019, \pasj, 71, 6. doi:10.1093/pasj/psy126
\bibitem[Hanaoka et al.(2020)]{2020PASJ...72....5H} Hanaoka, M., Kaneda, H., Suzuki, T., et al.\ 2020, \pasj, 72, 5. doi:10.1093/pasj/psz123
\bibitem[Hasegawa et al.(1994)]{1994ApJ...429L..77H} {Hasegawa, T., Sato, F., Whiteoak, J.~B., et al.\ 1994, \apjl, 429, L77. doi:10.1086/187417}
\bibitem[Hattori et al.(2016)]{2016PASJ...68...37H} Hattori, Y., Kaneda, H., Ishihara, D., et al.\ 2016, \pasj, 68, 37. doi:10.1093/pasj/psw028
\bibitem[Haworth et al.(2015)]{2015MNRAS.450...10H} Haworth, T.~J., Tasker, E.~J., Fukui, Y., et al.\ 2015, \mnras, 450, 10. doi:10.1093/mnras/stv639
\bibitem[Helfand et al.(2006)]{2006AJ....131.2525H} Helfand, D.~J., Becker, R.~H., White, R.~L., et al.\ 2006, \aj, 131, 2525. doi:10.1086/503253

\bibitem[Higuchi et al.(2014)]{2014AJ....147..141H} Higuchi, A.~E., Chibueze, J.~O., Habe, A., et al.\ 2014, \aj, 147, 141. doi:10.1088/0004-6256/147/6/141

\bibitem[Hirata et al.(2022)]{hirata2022} Hirata, Y., Handa, T., Murase. T., et al.\ 2022, in preparation.
\bibitem[Handa et al.(2006)]{2006JPhCS..54...42H} {Handa, T., Omodaka, T., Nagayama, T., et al.\ 2006, Journal of Physics Conference Series, 54, 42. doi:10.1088/1742-6596/54/1/007}
\bibitem[Ho \& Townes(1983)]{1983ARA&A..21..239H} Ho, P.~T.~P. \& Townes, C.~H.\ 1983, \araa, 21, 239.
doi:10.1146/annurev.aa.21.090183.001323
\bibitem[Hosokawa \& Inutsuka(2006)]{2006ApJ...646..240H} Hosokawa, T. \& Inutsuka, S.-. ichiro .\ 2006, \apj, 646, 240. doi:10.1086/504789
\bibitem[Hunter(2007)]{2007CSE.....9...90H} Hunter, J.~D.\ 2007, Computing in Science and Engineering, 9, 90
\bibitem[Ikeda et al.(2001)]{2001ASPC..238..522I} Ikeda, M., Nishiyama, K., Ohishi, M., et al.\ 2001, Astronomical Data Analysis Software and Systems X, 238, 522
\bibitem[Jackson et al.(2006)]{2006ApJS..163..145J} Jackson, J.~M., Rathborne, J.~M., Shah, R.~Y., et al.\ 2006, \apjs, 163, 145. doi:10.1086/500091
\bibitem[Kamazaki et al.(2012)]{2012PASJ...64...29K} Kamazaki, T., Okumura, S.~K., Chikada, Y., et al.\ 2012, \pasj, 64, 29
\bibitem[Kauffmann et al.(2008)]{2008A&A...487..993K} Kauffmann, J., Bertoldi, F., Bourke, T.~L., et al.\ 2008, \aap, 487, 993. doi:10.1051/0004-6361:200809481
\bibitem[Kawamura et al.(1998)]{1998ApJS..117..387K} {Kawamura, A., Onishi, T., Yonekura, Y., et al.\ 1998, \apjs, 117, 387. doi:10.1086/313119}
\bibitem[Kohno et al.(2018)]{2018PASJ...70S..50K} Kohno, M., Torii, K., Tachihara, K., et al.\ 2018, \pasj, 70, S50. doi:10.1093/pasj/psx137
\bibitem[Kohno et al.(2021a)]{2021PASJ...73S.129K} Kohno, M., Tachihara, K., Torii, K., et al.\ 2021a, \pasj, 73, S129. doi:10.1093/pasj/psaa015
\bibitem[Kohno et al.(2021b)]{2021PASJ...73S.338K} Kohno, M., Tachihara, K., Fujita, S., et al.\ 2021b, \pasj, 73, S338. doi:10.1093/pasj/psy109
\bibitem[Kohno et al.(2022)]{2022S255} Kohno, M., Omodaka, T., Handa, T., et al.\ 2022, \pasj, 74, 545. doi:10.1093/pasj/psac014
\bibitem[Koide et al.(2019)]{2019PASJ...71..113K} {Koide, N., Nakanishi, H., Sakai, N., et al.\ 2019, \pasj, 71, 113. doi:10.1093/pasj/psz101}

\bibitem[Kuno et al.(2011)]{Proc..2011} Kuno, N., et al. 2011, in Proc. 2011 XXXth URSI General Assembly
and Scientific Symposium (New York: IEEE), 3670 \footnote{\url{http://ieeexplore.ieee.org/xpl/articleDetails.jsp?arnumber=6051296}}
\bibitem[Kutner, \& Ulich(1981)]{1981ApJ...250..341K} Kutner, M.~L., \& Ulich, B.~L.\ 1981, \apj, 250, 341
\bibitem[Ladeyschikov et al.(2019)]{2019AJ....158..233L} Ladeyschikov, D.~A., Bayandina, O.~S., \& Sobolev, A.~M.\ 2019, \aj, 158, 233. doi:10.3847/1538-3881/ab4b4c
\bibitem[Li et al.(2019)]{2019MNRAS.487.1517L} Li, X., Esimbek, J., Zhou, J., et al.\ 2019, \mnras, 487, 1517. doi:10.1093/mnras/stz1269
\bibitem[Luisi et al.(2021)]{2021SciA....7.9511L} Luisi, M., Anderson, L.~D., Schneider, N., et al.\ 2021, Science Advances, 7, eabe9511. doi:10.1126/sciadv.abe9511
\bibitem[Mangum et al.(1992)]{1992ApJ...388..467M} Mangum, J.~G., Wootten, A., \& Mundy, L.~G.\ 1992, \apj, 388, 467. doi:10.1086/171167
\bibitem[Mangum \& Shirley(2015)]{2015PASP..127..266M} Mangum, J.~G. \& Shirley, Y.~L.\ 2015, \pasp, 127, 266. doi:10.1086/680323
\bibitem[Mauch et al.(2020)]{2020ApJ...888...61M} Mauch, T., Cotton, W.~D., Condon, J.~J., et al.\ 2020, \apj, 888, 61. doi:10.3847/1538-4357/ab5d2d
\bibitem[McGary \& Ho(2002)]{2002ApJ...577..757M} McGary, R.~S. \& Ho, P.~T.~P.\ 2002, \apj, 577, 757. doi:10.1086/342233
\bibitem[McGlynn et al.(1998)]{1998IAUS..179..465M} McGlynn, T., Scollick, K., \& White, N.\ 1998, New Horizons from Multi-Wavelength Sky Surveys, 179, 465
\bibitem[Minamidani et al.(2016)]{2016SPIE.9914E..1ZM} Minamidani, T., Nishimura, A., Miyamoto, Y., et al.\ 2016, \procspie, 9914, 99141Z. doi:10.1117/12.2232137
\bibitem[Morris et al.(1973)]{1973ApJ...186..501M} {Morris, M., Zuckerman, B., Palmer, P., et al.\ 1973, \apj, 186, 501. doi:10.1086/152515}
\bibitem[Murase et al.(2020)]{2020IAUS..345..353M} {Murase, T., Handa, T., Maebata, M., et al.\ 2020, Origins: From the Protosun to the First Steps of Life, 345, 353. doi:10.1017/S174392131900200X}
\bibitem[Murase et al.(2022)]{2021arXiv211113481M} Murase, T., Handa, T., Hirata, Y., et al.\ 2022, \mnras, 510, 1106. doi:10.1093/mnras/stab3472
\bibitem[Nagahama et al.(1998)]{1998AJ....116..336N} {Nagahama, T., Mizuno, A., Ogawa, H., et al.\ 1998, \aj, 116, 336. doi:10.1086/300392}
\bibitem[Nagayama et al.(2007)]{2007PASJ...59..869N} Nagayama, T., Omodaka, T., Handa, T., et al.\ 2007, \pasj, 59, 869. doi:10.1093/pasj/59.5.869
\bibitem[Nagayama et al.(2009)]{2009PASJ...61.1023N} Nagayama, T., Omodaka, T., Handa, T., et al.\ 2009, \pasj, 61, 1023. doi:10.1093/pasj/61.5.1023
\bibitem[Nakajima et al.(2019)]{2019PASJ...71S..17N} Nakajima, T., Inoue, H., Fujii, Y., et al.\ 2019, \pasj, 71, S17
\bibitem[Nakano et al.(2017)]{2017PASJ...69...16N} Nakano, M., Soejima, T., Chibueze, J.~O., et al.\ 2017, \pasj, 69, 16. doi:10.1093/pasj/psw120
\bibitem[Nakanishi et al.(2020)]{2020PASJ...72...43N} {Nakanishi, H., Fujita, S., Tachihara, K., et al.\ 2020, \pasj, 72, 43. doi:10.1093/pasj/psaa027}
\bibitem[Nishimura et al.(2018)]{2018PASJ...70S..42N} Nishimura, A., Minamidani, T., Umemoto, T., et al.\ 2018, \pasj, 70, S42
\bibitem[Nishimura et al.(2021)]{2021PASJ...73S.285N} {Nishimura, A., Fujita, S., Kohno, M., et al.\ 2021, \pasj, 73, S285. doi:10.1093/pasj/psaa083}
\bibitem[Ohama et al.(2018)]{2018PASJ...70S..47O} Ohama, A., Kohno, M., Fujita, S., et al.\ 2018, \pasj, 70, S47. doi:10.1093/pasj/psy012
\bibitem[Perez, \& Granger(2007)]{2007CSE.....9c..21P} Perez, F., \& Granger, B.~E.\ 2007, Computing in Science and Engineering, 9, 21
\bibitem[Pickett et al.(1998)]{1998JQSRT..60..883P} Pickett, H.~M., Poynter, R.~L., Cohen, E.~A., et al.\ 1998, \jqsrt, 60, 883. doi:10.1016/S0022-4073(98)00091-0
\bibitem[Purcell et al.(2012)]{2012MNRAS.426.1972P} Purcell, C.~R., Longmore, S.~N., Walsh, A.~J., et al.\ 2012, \mnras, 426, 1972. doi:10.1111/j.1365-2966.2012.21800.x
\bibitem[Palmeirim et al.(2017)]{2017A&A...605A..35P} {Palmeirim, P., Zavagno, A., Elia, D., et al.\ 2017, \aap, 605, A35. doi:10.1051/0004-6361/201629963}
\bibitem[Priestley \& Whitworth(2021)]{2021MNRAS.506..775P} Priestley, F.~D. \& Whitworth, A.~P.\ 2021, \mnras, 506, 775. doi:10.1093/mnras/stab1777
\bibitem[Robitaille et al.(2006)]{2006ApJS..167..256R} Robitaille, T.~P., Whitney, B.~A., Indebetouw, R., et al.\ 2006, \apjs, 167, 256. doi:10.1086/508424
\bibitem[Robitaille \& Bressert(2012)]{2012ascl.soft08017R} Robitaille, T., \& Bressert, E.\ 2012, APLpy: Astronomical Plotting Library in Python, ascl:1208.017
\bibitem[Sato et al.(2021)]{2021PASJ...73..568S} {Sato, K., Hasegawa, T., Umemoto, T., et al.\ 2021, \pasj, 73, 568. doi:10.1093/pasj/psab021}
\bibitem[Sault et al.(1995)]{1995ASPC...77..433S} {Sault, R.~J., Teuben, P.~J., \& Wright, M.~C.~H.\ 1995, Astronomical Data Analysis Software and Systems IV, 77, 433}
\bibitem[Sawada et al.(2008)]{2008PASJ...60..445S} Sawada, T., Ikeda, N., Sunada, K., et al.\ 2008, \pasj, 60, 445
\bibitem[Schuller et al.(2009)]{2009A&A...504..415S} Schuller, F., Menten, K.~M., Contreras, Y., et al.\ 2009, \aap, 504, 415. doi:10.1051/0004-6361/200811568
\bibitem[Sofue et al.(2019)]{2019PASJ...71S...1S} {Sofue, Y., Kohno, M., Torii, K., et al.\ 2019, \pasj, 71, S1. doi:10.1093/pasj/psy094}
\bibitem[Solomon et al.(1979)]{1979ApJ...232L..89S} {Solomon, P.~M., Scoville, N.~Z., \& Sanders, D.~B.\ 1979, \apjl, 232, L89. doi:10.1086/183042}
\bibitem[Tafalla et al.(2004)]{2004A&A...416..191T} Tafalla, M., Myers, P.~C., Caselli, P., et al.\ 2004, \aap, 416, 191. doi:10.1051/0004-6361:20031704
\bibitem[Takahira et al.(2014)]{2014ApJ...792...63T} Takahira, K., Tasker, E.~J., \& Habe, A.\ 2014, \apj, 792, 63. doi:10.1088/0004-637X/792/1/63
\bibitem[Takeba et al.(2022)]{Takeba2022} {Takeba, N., Handa, T., Murase, T., et al.\ 2023, IAU Proceeding, in press.}

\bibitem[Thompson et al.(2012)]{2012MNRAS.421..408T} {Thompson, M.~A., Urquhart, J.~S., Moore, T.~J.~T., et al.\ 2012, \mnras, 421, 408. doi:10.1111/j.1365-2966.2011.20315.x}
\bibitem[Torii et al.(2015)]{2015ApJ...806....7T} Torii, K., Hasegawa, K., Hattori, Y., et al.\ 2015, \apj, 806, 7. doi:10.1088/0004-637X/806/1/7
\bibitem[Torii et al.(2019)]{2019PASJ...71S...2T} Torii, K., Fujita, S., Nishimura, A., et al.\ 2019, \pasj, 71, S2. doi:10.1093/pasj/psz033
\bibitem[Toujima et al.(2011)]{2011PASJ...63.1259T} Toujima, H., Nagayama, T., Omodaka, T., et al.\ 2011, \pasj, 63, 1259. doi:10.1093/pasj/63.6.1259
\bibitem[Turner(1991)]{1991ApJS...76..617T} Turner, B.~E.\ 1991, \apjs, 76, 617. doi:10.1086/191577
\bibitem[Ulich, \& Haas(1976)]{1976ApJS...30..247U} Ulich, B.~L., \& Haas, R.~W.\ 1976, \apjs, 30, 247
\bibitem[Umemoto et al.(2017)]{2017PASJ...69...78U} Umemoto, T., Minamidani, T., Kuno, N., et al.\ 2017, \pasj, 69, 78. doi:10.1093/pasj/psx061
\bibitem[Ungerechts et al.(1986)]{1986A&A...157..207U} Ungerechts, H., Walmsley, C.~M., \& Winnewisser, G.\ 1986, \aap, 157, 207
\bibitem[Urquhart et al.(2011)]{2011MNRAS.418.1689U} Urquhart, J.~S., Morgan, L.~K., Figura, C.~C., et al.\ 2011, \mnras, 418, 1689.
doi:10.1111/j.1365-2966.2011.19594.x
\bibitem[Urquhart et al.(2014)]{2014A&A...568A..41U} Urquhart, J.~S., Csengeri, T., Wyrowski, F., et al.\ 2014, \aap, 568, A41. doi:10.1051/0004-6361/201424126
\bibitem[Urquhart et al.(2015)]{2015MNRAS.452.4029U} Urquhart, J.~S., Figura, C.~C., Moore, T.~J.~T., et al.\ 2015, \mnras, 452, 4029. doi:10.1093/mnras/stv1514
\bibitem[van der Walt et al.(2011)]{2011CSE....13b..22V} van der Walt, S., Colbert, S.~C., \& Varoquaux, G.\ 2011, Computing in Science and Engineering, 13, 22
\bibitem[Walmsley \& Ungerechts(1983)]{1983A&A...122..164W} Walmsley, C.~M. \& Ungerechts, H.\ 1983, \aap, 122, 164
\bibitem[Watson et al.(2008)]{2008ApJ...681.1341W} Watson, C., Povich, M.~S., Churchwell, E.~B., et al.\ 2008, \apj, 681, 1341. doi:10.1086/588005
\bibitem[Weaver et al.(1977)]{1977ApJ...218..377W} Weaver, R., McCray, R., Castor, J., et al.\ 1977, \apj, 218, 377. doi:10.1086/155692
\bibitem[Werner et al.(2004)]{2004ApJS..154....1W} Werner, M.~W., Roellig, T.~L., Low, F.~J., et al.\ 2004, \apjs, 154, 1. doi:10.1086/422992
\bibitem[White et al.(2005)]{2005AJ....130..586W} White, R.~L., Becker, R.~H., \& Helfand, D.~J.\ 2005, \aj, 130, 586. doi:10.1086/431249
\bibitem[Whitworth et al.(1994)]{1994MNRAS.268..291W} {Whitworth, A.~P., Bhattal, A.~S., Chapman, S.~J., et al.\ 1994, \mnras, 268, 291. doi:10.1093/mnras/268.1.291}
\bibitem[Wienen et al.(2012)]{2012A&A...544A.146W} Wienen, M., Wyrowski, F., Schuller, F., et al.\ 2012, \aap, 544, A146. doi:10.1051/0004-6361/201118107
\bibitem[Wienen et al.(2018)]{2018A&A...609A.125W} Wienen, M., Wyrowski, F., Menten, K.~M., et al.\ 2018, \aap, 609, A125. doi:10.1051/0004-6361/201526384
\bibitem[Wilson et al.(2013)]{2013tra..book.....W} Wilson, T.~L., Rohlfs, K., \& H{\"u}ttemeister, S.\ 2013, Tools of Radio Astronomy; Astronomy and Astrophysics Library. ISBN 978-3-642-39949-7. Springer-Verlag Berlin Heidelberg
\bibitem[Xu et al.(2019)]{2019RAA....19..183X} Xu, J.-L., Stutzki, J., Wu, Y., et al.\ 2019, Research in Astronomy and Astrophysics, 19, 183. doi:10.1088/1674-4527/19/12/183
\bibitem[Xu \& Ju(2014)]{2014A&A...569A..36X} Xu, J.-L. \& Ju, B.-G.\ 2014, \aap, 569, A36. doi:10.1051/0004-6361/201423952
\bibitem[Yamada et al.(2021)]{2021PASJ...73..880Y} {Yamada, R.~I., Enokiya, R., Sano, H., et al.\ 2021, \pasj, 73, 880. doi:10.1093/pasj/psab050}
\bibitem[Yan et al.(2016)]{2016AJ....152..117Y} Yan, Q.-. zeng ., Xu, Y., Zhang, B., et al.\ 2016, \aj, 152, 117. doi:10.3847/0004-6256/152/5/117
\bibitem[Zavagno et al.(2006)]{2006A&A...446..171Z} Zavagno, A., Deharveng, L., Comer{\'o}n, F., et al.\ 2006, \aap, 446, 171. doi:10.1051/0004-6361:20053952
\bibitem[Zavagno et al.(2007)]{2007A&A...472..835Z} Zavagno, A., Pomar{\`e}s, M., Deharveng, L., et al.\ 2007, \aap, 472, 835. doi:10.1051/0004-6361:20077474
\bibitem[Zavagno et al.(2010)]{2010A&A...518L.101Z} Zavagno, A., Anderson, L.~D., Russeil, D., et al.\ 2010, \aap, 518, L101. doi:10.1051/0004-6361/201014587
\end{thebibliography}
\end{document}